\documentclass[12pt]{article}
\usepackage{amsmath,amssymb,amsthm,amsxtra,overpic,bbm,bm,epsfig,sub
figure,cite}
\usepackage{mathrsfs}
\usepackage{graphicx}
\usepackage{color}
\usepackage{comment}
\usepackage{epstopdf}
\usepackage{float}
\usepackage{slashed,stmaryrd}

\newcommand{\RNum}[1]{\uppercase\expandafter{\romannumeral #1\relax}}
\def\thefootnote{\fnsymbol{footnote}}

\begin{document}
\begin{center}
	{\Large\bf Radiative corrections to the lepton flavor mixing\\
		in dense matter}
\end{center}

\begin{center}
{\bf Jing-yu Zhu$^{\dagger}$} \footnote{E-mail:
		zhujingyu@sjtu.edu.cn}
	\\
	{\small $^{\dagger}$School of Physics and Astronomy and Tsung-Dao Lee Institute, 
		Shanghai Jiao Tong University, Shanghai 200240, China \\}
\end{center}

\vspace{1.5cm}
\begin{abstract}
One-loop radiative corrections will lead to a small difference between the matter potentials developed
by $\nu_\mu^{}$ and $\nu_{\tau}$ when they travel in a medium. 
By including such radiative corrections, we derive the exact expressions of the corresponding effective mass-squared differences and the
moduli square of the lepton flavor mixing matrix elements $|\widetilde U_{\alpha i}^{}|^2$ (for $\alpha=e,\mu,\tau $ and $i=1,2,3$) in matter in the standard three-flavor mixing scheme and focus on their asymptotic behaviors when the matter density is very big  (i.e., the matter effect parameter $A\equiv 2\sqrt{2} G_{\rm F}^{} N_e^{}E$ is very big).  Different from the non-trivial fixed value of $|\widetilde U_{\alpha i}^{}|^2$ in the $A\to \infty$ limit in the case without radiative corrections, we get $|\widetilde U_{\alpha i}^{}|^2=0~{\rm or}~1$ under this extreme condition. The radiative corrections can significantly affect the lepton flavor mixing in dense matter,
which are numerically and analytically discussed in detail.
Furthermore, we also extend the discussion to the $(3+1)$ active-sterile neutrino mixing scheme.
\end{abstract}
\begin{flushleft}
	\hspace{0.8cm} PACS number(s): 14.60.Pq, 25.30.Pt
\end{flushleft}

\def\thefootnote{\arabic{footnote}}
\setcounter{footnote}{0}

\newpage
\vspace{0.2cm}
\section{Introduction}
In 1978, Wolfenstein firstly pointed out that when neutrinos travel in matter, the coherent forward scattering of them with electrons and nucleons must be considered and the induced matter potentials will change the neutrino oscillation behaviors \cite{Wolfenstein:1977ue}.  
In 1985,  Mikheev and Smirnov put forward that  the effective mixing angle can be significantly amplified in matter (such as inside the sun) even if the corresponding mixing angle in vacuum is small. This is the wellknown Mikheev-Smirnov-Wolfenstein (MSW) effects, which successfully explain the flavor conversion behaviors of solar neutrinos in the sun \cite{Mikheev:1986wj}. Such matter effects have been proved very important in a number of reactor, solar, atmospheric, accelerator neutrino oscillation experiments aiming to
accurately extract the intrinsic neutrino oscillation parameters in vacuum \cite{Tanabashi:2018oca}. 
A lot of efforts have been made to make the neutrino oscillation probabilities in matter more
intuitive \cite{Freund:2001pn, Akhmedov:2004ny, He:2013rba,Xu:2015kma, He:2016dco,Xing:2016ymg,Denton:2016wmg, Li:2016txk,Li:2016pzm,Ioannisian:2018qwl,Li:2018jgd,Denton:2018fex, Petcov:2018zka,Wang:2019dal}. The language of renormalization-group equation was also introduced to describe
the effective neutrino masses and flavor mixing parameters in matter \cite{Zhou:2016luk, Chiu:2017ckv,Xing:2018lob,Wang:2019yfp}. In this paper, we mainly
focus on the neutrino flavor mixing in very dense matter, which has
been discussed in Refs.  \cite{Blennow:2004js, Huang:2018ufu, Xing:2019owb,Luo:2019efb}.  
We further include the radiative corrections in this connection.

In the standard three-flavor mixing scheme, the Hamiltonian responsible for the propagation of neutrinos
in matter can be expressed as
\begin{eqnarray}
{\cal H}^{}_{\rm m} = \frac{1}{2E} U \left(\begin{matrix} m^2_1 & 0 & 0 \\ 0
& m^2_2 & 0 \\ 0 & 0 & m^2_3 \end{matrix}\right) U^\dagger +
\left(\begin{matrix} V^{}_{e}  & 0 & 0 \\ 0 &
V^{}_{\mu} & 0 \\
0 & 0 & V^{}_{\tau} \end{matrix}\right)
\equiv \frac{1}{2E}\widetilde U \left(\begin{matrix} \widetilde{m}^2_1 & 0 & 0 \\ 0
& \widetilde{m}^2_2 & 0 \\ 0 & 0 & \widetilde{m}^2_3 \end{matrix}\right)
\widetilde U^\dagger \; ,
%     (1)
\end{eqnarray}
where $E$ is the neutrino beam energy, $m_i^{}$ (for $i=1,2,3$) and $U$ stand respectively for neutrino masses and Pontecorvo-Maki-Nakagawa-Sakata (PMNS)  lepton flavor mixing matrix,  $\widetilde m_i^{}$ (for $i=1,2,3$) and $\widetilde U$ denote effective neutrino masses and PMNS matrix in matter, respectively, and $V_{\alpha}^{}~(\alpha=e,\mu,\tau)$ represent the matter potentials arsing from 
charged- and neutral-current coherent forward scattering of $\nu_{\alpha}^{}$ with electrons, protons and neutrons in matter. 
Considering the one-loop radiative corrections to $V_{\alpha}$, we have \cite{Botella:1986wy}
\begin{eqnarray}
V_{e}^{} - V_{\mu}^{}&=& \sqrt{2} G_{\rm F}^{} N_e^{} \;,\nonumber\\
V_{\tau}^{} - V_{\mu}^{}&=& -\frac{ G_{\rm F}^{}}{\sqrt{2}} \cdot \frac{3\alpha}{2 \pi \sin^2 \theta_{W}^{}} \cdot \frac{m_{\tau}^2}{m_W^{2}} \left[\left(N_p^{} + N_n^{}\right) {\rm ln} \left(\frac{m_{\tau}^2}{m_W^2}\right)+ \left(N_p+\frac{2}{3} N_n^{}\right)\right] \;,
%     (2)
\end{eqnarray}
where $G_{\rm F}^{}$ is the Fermi coupling constant, $\alpha$ is the fine-structure constant; $N_e^{}$, $N_p^{}$ and $N_n^{}$ denote the number density of electrons, protons and neutrons in matter, respectively;  $m_\tau^{}$ and $m_W^{}$  are the masses of $\tau$ lepton and $W$ boson, respectively; and $\sin^2
\theta_{W}^{} \equiv 1- m_W^2/m_Z^2$ with $m_Z^{}$ being the $Z$ boson mass.
To be more intuitive, Eq. (1) can be rewritten as
\begin{eqnarray}
{\cal H}^{}_{\rm m} = \frac{1}{2E} \left[U \left(\begin{matrix} 0 & 0 & 0 \\ 0
& \Delta^{}_{21} & 0 \\ 0 & 0 & \Delta^{}_{31} \end{matrix}\right) U^\dagger +
\left(\begin{matrix} A & 0 & 0 \\ 0 & 0 & 0 \\ 0 & 0 & A\epsilon \end{matrix}\right)\right]
\equiv \frac{1}{2E} \left[\widetilde{U} \left(\begin{matrix} 0 & 0 & 0 \\ 0
& \widetilde{\Delta}^{}_{21} & 0 \\ 0 & 0 & \widetilde{\Delta}^{}_{31}
\end{matrix}\right) \widetilde{U}^\dagger + BI\right] \; ,
%     (3)
\end{eqnarray}
where $\Delta_{ij} \equiv m_i^2 - m_j^2$,  $\widetilde\Delta_{ij} \equiv \widetilde m_i^2 - \widetilde m_j^2$, $A= 2 \sqrt{2} G_{\rm F}^{}N_e^{}E $, $B = \widetilde{m}^2_1 - m^2_1 -  2E V^{}_{\mu}$,
and $I$ denotes a $3\times 3$ identity matrix.
According to Eq. (2) and assuming $N_e^{}=N_{p}^{}=N_{n}$,  $\epsilon\simeq 5 \times 10^{-5}$ is a small quantity but matters a lot in dense matter (i.e., A is very big).

On  the other hand, given the implications of extra light sterile neutrinos in short-baseline neutrino oscillation
experiments \cite{Boser:2019rta}, we extend our discussion to the scheme of $(3+1)$ flavor mixing with one more
sterile neutrino $\nu_s^{}$. The corresponding Hamiltonian describing the propagation of neutrinos in a medium turns out to be
\begin{eqnarray}
{\cal H}^{s}_{\rm m} &=& \frac{1}{2E} \left[V \left(\begin{matrix} 0 & 0 & 0&0 \\ 0
& \Delta^{}_{21}&0 & 0 \\ 0 & 0 & \Delta^{}_{31} &0 \\0&0&0&\Delta_{41}\end{matrix}\right) V^\dagger +
\left(\begin{matrix} A & 0 & 0 &0\\ 0 & 0 & 0&0 \\ 0 & 0 & A\epsilon&0\\0&0&0&A^{\prime} \end{matrix}\right)\right] \nonumber\\
&=&\frac{1}{2E} \left[\widetilde{V} \left(\begin{matrix} 0 & 0 & 0&0 \\ 0
& \widetilde{\Delta}^{}_{21} & 0 &0\\ 0 & 0 & \widetilde{\Delta}^{}_{31}&0\\0&0&0&
\widetilde \Delta_{41}^{}
\end{matrix}\right) \widetilde{V}^\dagger + B I\right] \; ,
%     (4)
\end{eqnarray}
where $V$ and $\widetilde V$ denote the $4\times 4$ lepton flavor mixing matrix in
vacuum and matter, respectively, $A^{\prime}=-2 E V_{\mu}^{}=\sqrt{2} G_{\rm F}^{} N_n^{} E$, $\Delta_{41}^{} =m_4^2 -m_1^2$, and $\widetilde \Delta_{41}^{} =\widetilde m_4^2 - \widetilde m_1^2$ with $m_4^{}$ and $\widetilde m_4^{}$ being the sterile neutrino mass in vacuum and
matter, respectively. The existence of an extra sterile neutrino can make the neutrino flavor mixing in
dense matter very different from the standard three-flavor mixing scheme.

The remaining parts of this paper are organized as follows. In section 2,  we include the 
radiative corrections and derive the
corresponding expressions of $\widetilde \Delta_{ij}^{}$,
$|\widetilde U_{\alpha i}^{}|^2$ and $\widetilde U_{\alpha i}^{}\widetilde U_{\beta i}^{*}$ in matter in the standard three-flavor mixing scheme. The asymptotic behaviors of $\widetilde \Delta_{ij}^{}$ and $|\widetilde U_{\alpha i}^{}|^2$ and neutrino oscillations in dense matter are analytically and numerically investigated in detail.
In section 3, we extend our discussion to the $(3+1)$ flavor mixing scheme.
Section 4 is devoted to a brief summary.
%%%%%%%%%%%%%%%%              section      2   %%%%%%%%%%%%%%%%%%%%%%%%%%%
\section{The standard three-flavor mixing scheme}
In the standard three-flavor mixing scheme, the exact formulas of $ \widetilde{\Delta}^{}_{ij}$ without radiative corrections have been given in Refs. \cite{Barger:1980tf, Zaglauer:1988gz,Xing:2000gg}.
Considering radiative correction effects, we derive the eigenvalues of ${\cal H}_{\rm m}^{}$ in Eq. (3) and express the two independent effective mass-squared differences $\widetilde\Delta_{ij}^{}$ (for $ij=21,31$) as
\begin{eqnarray}
&& \widetilde{\Delta}^{}_{21} =
\frac{2}{3} \sqrt{x^2 -3 y} \sqrt{3 \left(1 - z^2\right)} \;\; ,
\nonumber \\
&& \widetilde{\Delta}^{}_{31} =
\frac{1}{3} \sqrt{x^2 -3 y} \left[3z + \sqrt{3 \left(1 - z^2\right)}
\right] \; ,
%     (5)
\end{eqnarray}
for the case of normal mass ordering (NMO) with $m^{}_1 < m^{}_2 < m^{}_3$; or
\begin{eqnarray}
&& \widetilde{\Delta}^{}_{21} =
\frac{1}{3} \sqrt{x^2 - 3 y} \left[3z - \sqrt{3 \left(1 - z^2\right)}\right] \; ,
\nonumber \\
&& \widetilde{\Delta}^{}_{31} = -\frac{2}{3} \sqrt{x^2 - 3 y} \sqrt{3 \left(1 -
	z^2\right)} \;\; ,
%     (6)
\end{eqnarray}
for the case of inverted mass ordering (IMO) case with
$m^{}_3 < m^{}_1 < m^{}_2$, where
\begin{eqnarray}
x & = & \Delta^{}_{21} + \Delta^{}_{31} + A (1+\epsilon)\; ,
\nonumber \\
y & = & \Delta^{}_{21} \Delta^{}_{31} + A \left\{\Delta^{}_{21} \left[1 -
|U^{}_{e2}|^2 +\epsilon\left(1-|U^{}_{\tau 2}|^2\right)\right]  \right.\nonumber\\
&&\left. +\Delta^{}_{31}\left[ \left(1- |U^{}_{e3}|^2\right)+\epsilon\left(1-|U^{}_{\tau 3}|^2\right)\right]\right\}+A^2\epsilon\; ,
\nonumber \\
z & = & \cos\left[\frac{1}{3}\arccos\frac{2 x^3 - 9 xy + 27 d}{2 \sqrt{\left(x^2 - 3y\right)^3}}\right] 
%     (5)
\end{eqnarray}
with $d=A\Delta_{21}^{}\Delta_{31}^{}(|U_{e1}^{}|^2+ \epsilon |U_{\tau 1}^{}|^2)+A^2\epsilon(|U_{\mu2}^{}|^2\Delta_{21}^{} +|U_{\mu3}^{}|^2\Delta_{31}^{})$.
Taking the trace of ${\cal H}_{\rm m}^{}$ yields
\begin{eqnarray}
B = \frac{1}{3} \left[\Delta_{21}^{} + \Delta_{31}^{} + A\left(1 + \epsilon\right) -
\widetilde \Delta_{21}^{} - \widetilde \Delta_{31}^{}\right] \; .
%     (8)
\end{eqnarray}
The unitarity conditions of $\widetilde U$ and the sum rules derived from  ${\cal H}_{\rm m}^{}$ and ${\cal H}_{\rm m}^{2}$,
constitute a set of linear equations of $\widetilde U_{\alpha i}^{}\widetilde U_{\beta i}^{*}$  (for $\alpha,\beta=e,\mu,\tau$ and $i=1,2,3$):
\begin{eqnarray}
&&\sum_i^{} \widetilde U_{\alpha i}^{} \widetilde U_{\beta i}^{*}= \delta_{\alpha\beta} \; ,
\nonumber \\
&&\sum_i^{} \widetilde \Delta_{i1}^{}\widetilde U_{\alpha i}^{} \widetilde U_{\beta i}^{*}= 
\sum_i^{}  \Delta_{i1}^{}U_{\alpha i}^{} U_{\beta i}^{*} + {\cal A}_{\alpha\beta}^{} - B \delta_{\alpha\beta}
\;, \nonumber \\
&&\sum_i^{} \widetilde \Delta_{i1}^{}\left(\widetilde \Delta_{i1}^{} + 2 B\right)\widetilde U_{\alpha i}^{} \widetilde U_{\beta i}^{*}= 
\sum_i^{}  \Delta_{i1}^{}\left(\Delta_{i1}^{} + {\cal A}_{\alpha\alpha} + {\cal A}_{\beta\beta}\right)U_{\alpha i}^{} U_{\beta i}^{*}  + {\cal A}_{\alpha\beta}^2- B^2 \delta_{\alpha\beta} \;,
%      (9)
\end{eqnarray}
where ${\cal A}_{\alpha\beta}^{}$ stand for the $(\alpha,~\beta)$ element of the matter potential matrix ${\cal A}\equiv{\rm Diag}\{A, 0, A\epsilon\}$. Taking $\alpha=\beta$ and solving Eq. (9),
we obtain  
\begin{eqnarray}
|\widetilde U_{\alpha i}^{}|^2 &=& \frac{1}{\prod \limits^{}_{k\neq i}\widetilde \Delta_{ik}^{}} \sum_j^{} \left(F_{\alpha}^{ij}  |U_{\alpha j}^{}|^2 \right)
%       (10)
\end{eqnarray}
with
\begin{eqnarray}
F_{\alpha}^{ij} &=& \prod \limits^{}_{k\neq i} \left( \Delta_{j1}^{} - \widetilde\Delta_{k1}^{} - B+{\cal A}_{\alpha\alpha}^{}\right) \;,
%       (11)
\end{eqnarray}
where $\alpha=e,\mu,\tau$ and  $i,j,k=1,2,3$.
Similarly, in the case of $\alpha \neq\beta$, $\widetilde U_{\alpha i}^{} \widetilde U_{\beta i}^{*}$ can be derived
from Eq. (9):
\begin{eqnarray}
\widetilde U_{\alpha i}^{} \widetilde U_{\beta i}^{*}&=& \frac{1}{\prod \limits^{}_{k\neq i}\widetilde \Delta_{ik}^{}} \sum_{m= 1,2}^{} \left(F_{\alpha\beta}^{im}  U_{\alpha m}^{} U_{\beta m}^{*}\right)
%      (12)
\end{eqnarray}
with
\begin{eqnarray}
F_{\alpha\beta}^{im} &=& \Delta_{3m}^{}\left(\Delta_{n1} - \widetilde \Delta_{i1} - B + {\cal A}_{\gamma\gamma}\right) \;,
%       (13)
\end{eqnarray}
where $(\alpha,\beta,\gamma)$ run over $(e, \mu,\tau)$ and $n\neq m=1,2$. 
Note that $\widetilde U_{\alpha 1}^{} \widetilde U_{\beta 1}^{*}$,  $\widetilde U_{\alpha 2}^{} \widetilde U_{\beta 2}^{*}$ and  $\widetilde U_{\alpha 3}^{} \widetilde U_{\beta 3}^{*}$ for $\alpha\neq\beta$ constitute
the effective Dirac leptonic unitarity triangle in the complex plane.
From Eq. (12), it is straightforward to
check that the Naumov relation $\widetilde {\cal J}\widetilde \Delta_{21}^{}\widetilde \Delta_{31}^{}\widetilde \Delta_{32}^{} = {\cal J}\Delta_{21}^{}\Delta_{31}^{} \Delta_{32}^{}$  \cite{Naumov:1991ju} still holds, where ${\cal J}$ and $\widetilde{\cal J}$ are the Jarlskog invariants \cite{Jarlskog:1985ht} in vacuum and in matter, respectively,
\begin{eqnarray}
{\rm Im} (U^{}_{\alpha i} U^{}_{\beta j} U^*_{\alpha j} U^*_{\beta i})
& =&  {\cal J} \sum_\gamma \varepsilon^{}_{\alpha \beta \gamma}
\sum_k \varepsilon^{}_{ijk} \; , \hspace{0.8cm}
\nonumber \\
{\rm Im} (\widetilde{U}^{}_{\alpha i} \widetilde{U}^{}_{\beta j}
\widetilde{U}^*_{\alpha j} \widetilde{U}^*_{\beta i})
& =&  \widetilde{\cal J} \sum_\gamma \varepsilon^{}_{\alpha \beta \gamma}
\sum_k \varepsilon^{}_{ijk} \; ,
%     (14)
\end{eqnarray}
with $\varepsilon^{}_{\alpha \beta \gamma}$ and $\varepsilon^{}_{ijk}$ being
three-dimension Levi-Civita symbols.
The only difference due to the radiative corrections in the exact formulas of $\widetilde\Delta_{ij}^{}$,
$|\widetilde U_{\alpha i}^{}|^2$ and $\widetilde U_{\alpha i}^{} \widetilde U_{\beta i}^{*}$ above is the appearance of the term ${\cal A}_{\tau\tau}=A\epsilon$. By setting $\epsilon=0$,
one can turn off the radiative corrections and get the corresponding expressions of
$\widetilde\Delta_{ij}^{}$, $|\widetilde U_{\alpha i}^{}|^2$ and $\widetilde U_{\alpha i}^{} \widetilde U_{\beta i}^{*}$ in the previous literature  \cite{Xing:2000gg,Xing:2003ez,Zhang:2004hf,Xing:2005gk}.
With the help of Eqs. (5), (6) and (12), we can directly write out the probabilities of
the $\nu_{\alpha}^{} \to \nu_{\beta}^{}$ (for $\alpha,\beta=e,\mu,\tau$) oscillations in matter
\begin{eqnarray}
\widetilde P_{\alpha\beta}^{} &=& \delta_{\alpha\beta} - \sum_{i<j}^{} {\rm Re} \left( \widetilde U_{\alpha i}^{} \widetilde U_{\beta i}^{*} \widetilde U_{\alpha j}^{*} \widetilde U_{\beta j}^{}\right) 
\sin^2 \left(\frac{\widetilde\Delta_{ji}^{} L}{4 E}\right) \nonumber\\
&& + \sum_{i<j}^{} {\rm Im} \left( \widetilde U_{\alpha i}^{} \widetilde U_{\beta i}^{*} \widetilde U_{\alpha j}^{*} \widetilde U_{\beta j}^{}\right) 
\sin \left(\frac{\widetilde\Delta_{ji}^{} L}{2 E}\right)  \;,
% (15)
\end{eqnarray}
where $\alpha ,\beta=e,\mu,\tau$; $i,j=1,2,3$; and $L$ is the neutrino oscillation length.
Note that the results in Eqs. (5)--(15) are only valid for a neutrino beam. When it comes to an antineutrino beam, 
we need to do the replacements $U^{} \to {U}^{*}$ and $A\to -A$.  According to the exact
expressions of $\widetilde\Delta_{ij}^{}$,
$|\widetilde U_{\alpha i}^{}|^2$ and $\widetilde U_{\alpha i}^{} \widetilde U_{\beta i}^{*}$ in Eqs. (5), (6)
(10) and (12), we study the neutrino flavor mixing in dense matter in the
standard three-flavor mixing scheme. Both neutrinos and antineutrinos with the normal or inverted mass ordering (i.e., cases (NMO, $\nu$),  (IMO, $\nu$),  (NMO, $\overline\nu$) and (IMO, $\overline\nu$)) will be considered separately. Numerically, we take the standard parametrization of $U$,
\begin{eqnarray}
{U} = \left(\begin{matrix}
{c}^{}_{12} {c}^{}_{13} &{s}^{}_{12}
{c}^{}_{13} & {s}^{}_{13} e^{-{\rm i} {\delta}}
\cr -{s}^{}_{12} {c}^{}_{23} - {c}^{}_{12}
{s}^{}_{13} {s}^{}_{23} e^{{\rm i} {\delta}} &
{c}^{}_{12} {c}^{}_{23} -
{s}^{}_{12} {s}^{}_{13} {s}^{}_{23}
e^{{\rm i} {\delta}} & {c}^{}_{13}
{s}^{}_{23} \cr {s}^{}_{12} {s}^{}_{23} -
{c}^{}_{12}{s}^{}_{13} {c}^{}_{23}
e^{{\rm i} {\delta}} & -{c}^{}_{12} {s}^{}_{23}
- {s}^{}_{12} {s}^{}_{13} {c}^{}_{23}
e^{{\rm i} {\delta}} & {c}^{}_{13} {c}^{}_{23} \cr
\end{matrix} \right) \; , \hspace{0.3cm} 
%     (16)
\end{eqnarray}
and input the best-fit values of ($\theta_{12}^{},\theta_{13}^{},\theta_{23}^{},\delta,\Delta_{21}^{}, \Delta_{31}^{}$) in Ref. \cite{Esteban:2018azc}:
\begin{itemize}
	\item NMO: $\theta_{12}^{}=33.82^{\circ}$, $\theta_{13}^{}=8.60^{\circ}$, $\theta_{23}^{}=48.6^{\circ}$,
	$\delta=221^{\circ}$, $\Delta_{21}^{} = 7.39 \times 10^{-5}~ {\rm eV}^2$ and $\Delta_{31}^{} = 2.528 \times 10^{-3}~ {\rm eV}^2$; 
	\item IMO: $\theta_{12}^{}=33.82^{\circ}$, $\theta_{13}^{}=8.64^{\circ}$, $\theta_{23}^{}=48.8^{\circ}$,
	$\delta=282^{\circ}$, $\Delta_{21}^{} = 7.39 \times 10^{-5}~ {\rm eV}^2$ and $\Delta_{31}^{} = -2.436 \times 10^{-3}~ {\rm eV}^2$.
\end{itemize}
Analytically, we treat $\Delta_{21}^{}/A$, $\Delta_{31}^{}/A$ and $\epsilon$
as small quantities and make perturbative expansions of $\widetilde\Delta_{ij}^{}$ and $|\widetilde U_{\alpha i}^{}|^2$.
 Thus the analytical approximations in this section only apply to the range $A\gg\Delta_{31}^{}$.
\subsection{ (NMO, $\nu$)}
Let us first consider the case of a neutrino beam with normal mass ordering. 
The corresponding evolution of $\widetilde\Delta_{ij}^{}$ and $|\widetilde U_{\alpha i}^{}|^2$ with the matter effect parameter $A$ are illustrated in the upper left panel of Fig. 1 and  Fig. 2, respectively.
%%%%%%%%%%%%%%%%%% figure 1 %%%%%%%%%%%%%%%%%%%%%%%
\begin{figure}[h]
	\centering{
		\includegraphics[]{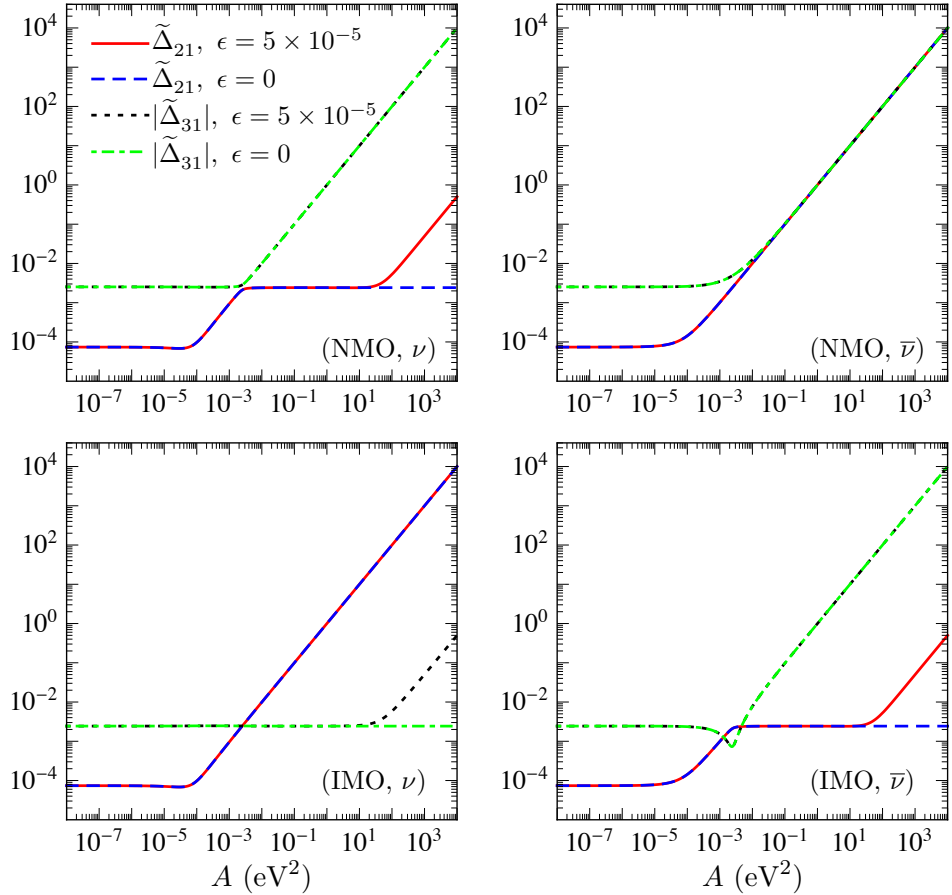}}
	\caption{In the standard three-flavor mixing scheme, the illustration of  how the effective neutrino mass-squared differences $\widetilde \Delta_{21}^{}$ and $|\widetilde \Delta_{31}^{}|$ evolve with the matter effect parameter $A$ in the
		cases with or without radiative corrections, where the best-fit values of 
		$(\theta_{12}^{}, \theta_{13}^{}, \theta_{23}^{}, \delta,\Delta_{21}^{}, \Delta_{31}^{} )$ in Ref. \cite{Esteban:2018azc} have been input.
	}
\end{figure}
%%%%%%%%%%%%%%%%%%%%%%%%%%%%%%%%%%%%%%%%%%%%%%%%%%%
%%%%%%%%%%%%%%%%%% figure 2 %%%%%%%%%%%%%%%%%%%%%%%
\begin{figure}[h]
	\centering{
		\includegraphics[width=16cm]{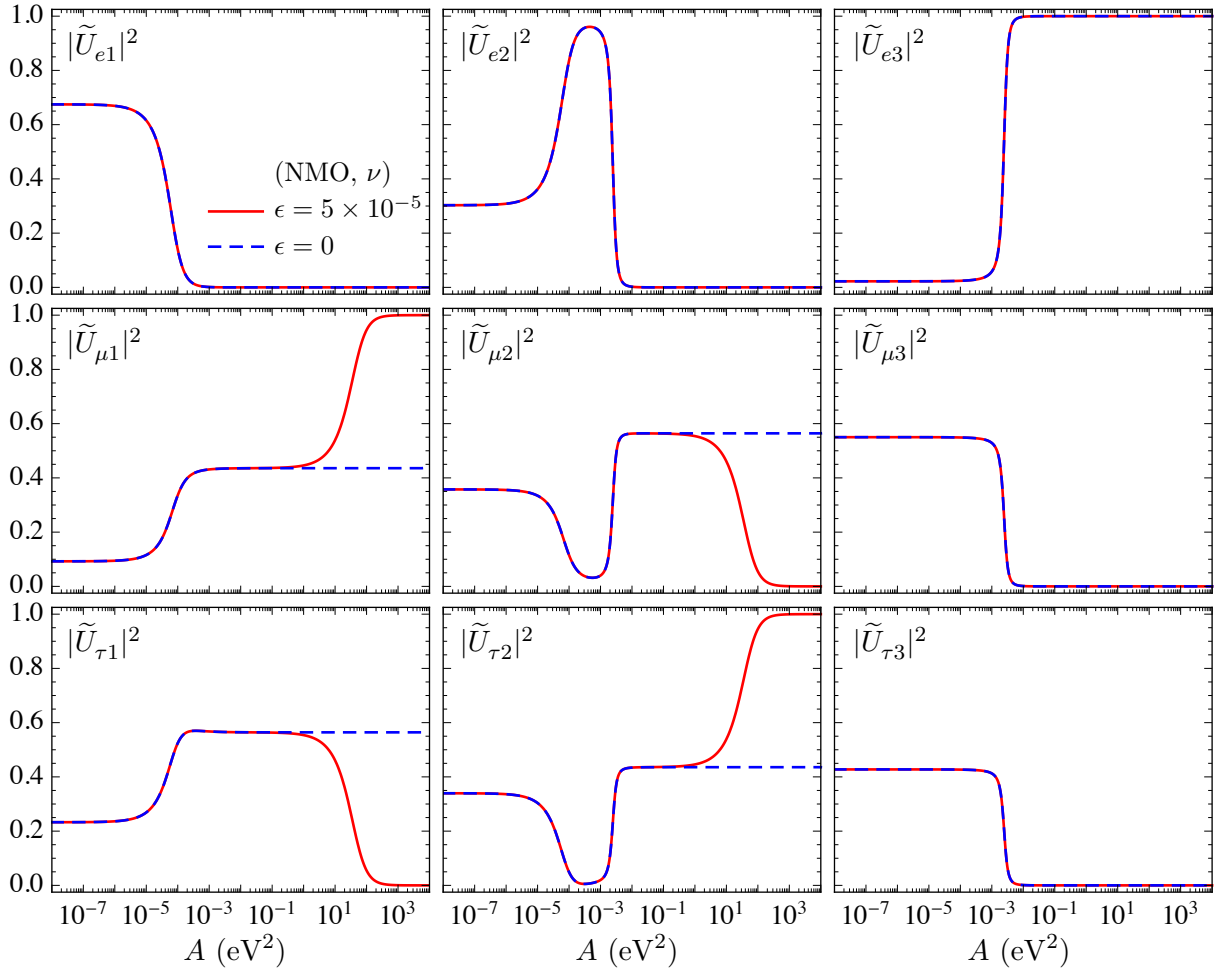}}
	\caption{In the standard three-flavor mixing scheme, the illustration of  how $|\widetilde  U_{\alpha i}^{}|^2$ (for $\alpha=e, \mu, \tau~{\rm and}~ i=1,2,3$) evolve with the matter effect parameter $A$ in the
		case $({\rm NMO},\nu)$ with or without radiative corrections, where the best-fit values of 
		$(\theta_{12}^{}, \theta_{13}^{}, \theta_{23}^{}, \delta,\Delta_{21}^{}, \Delta_{31}^{} )$ in Ref. \cite{Esteban:2018azc} have been input.}
\end{figure}
%%%%%%%%%%%%%%%%%%%%%%%%%%%%%%%%%%%%%%%%%%%%%%%%%%%
\noindent We find that the radiative corrections may significantly affect the values of $\widetilde\Delta_{ij}^{}$ and $|\widetilde U_{\alpha i}^{}|^2$ only if $A$ is big enough (for example, $A>1~{\rm eV}^2$). This can be revealed more clearly by expanding the exact expressions of $\widetilde \Delta_{ij}^{}$ and $|\widetilde U_{\alpha i}^{}|^2$ in terms of $\Delta_{21}^{}/A$, $\Delta_{31}^{}/A$ and $\epsilon$. Only keeping the
first order of these quantities, we simplify $\widetilde \Delta_{ij}^{}$ in Eq. (5) as
\begin{eqnarray}
 \widetilde{\Delta}^{}_{21} &\simeq& \xi\; ,
\nonumber \\
\widetilde{\Delta}^{}_{31} &\simeq& A-\frac{1}{2}\left[
A \epsilon+ (1 - 3 |U_{e2}^{}|^2) \Delta_{21}^{}+ (1 - 3 |U_{e3}^{}|^2) \Delta_{31}^{} -\xi
\right]\; , 
%     (17)
\end{eqnarray}
with
\begin{eqnarray}
\xi&=&\big\{
\left[\left( 1 - |U_{e2}^{}|^2  \right)\Delta_{21}^{}
+\left( 1 - |U_{e3}^{}|^2 \right) \Delta_{31}^{}+ A\epsilon\right]^2
-4 A \epsilon\left( |U_{\mu 2}^{}|^2 \Delta_{21} + |U_{\mu 3}^{}|^2 \Delta_{31} \right) \nonumber\\
&&- 4 |U_{e1}^{}|^2 \Delta_{21}^{} \Delta_{31}^{}
\big\}^{1/2} \;.
%              (18)
\end{eqnarray}
According to Eq. (17), it becomes clear that if $A$ is big enough, $\widetilde\Delta_{21}^{}$ will increase with
$A$ instead of taking a fixed value in the case without radiative corrections. The reason why the radiation corrections to $\widetilde\Delta_{31}^{}$ are not significant is just that
the much smaller $A\epsilon$ term appears in the next-to-leading order with the leading order being $A$.
By performing perturbative expansions of Eq. (10)  in terms of $\Delta_{21}^{}/A$, $\Delta_{31}^{}/A$ and $\epsilon$ and only keeping the leading order, $|\widetilde U_{\alpha i}^{}|^2$ are approximately expressed as
\begin{eqnarray}
|\widetilde{U}_{e1}^{}|^2  & \simeq & |\widetilde{U}_{e2}^{}|^2 \simeq
|\widetilde{U}_{\mu 3}^{}|^2 \simeq |\widetilde{U}_{\tau 3}^{}|^2 \simeq 0 \; ,
\quad |\widetilde{U}_{e3}^{}|^2 \simeq 1 \; ,
\nonumber \\
|\widetilde{U}_{\mu 1}^{}|^2  & \simeq & |\widetilde{U}_{\tau 2}^{}|^2 \simeq
\frac{1}{2} +\frac{ A \epsilon+ \Delta_{21}^{} \left(
	|U_{\tau 2}^{}|^2 - |U_{\mu 2}^{}|^2 \right) +
	\Delta_{31}^{} \left(|U_{\tau 3}^{}|^2 - |U_{\mu 3}^{}|^2
	\right)}{2 \xi} \; ,
\nonumber\\
|\widetilde{U}_{\mu 2}^{}|^2  & \simeq&  |\widetilde{U}_{\tau 1}^{}|^2 \simeq
\frac{1}{2} - \frac{A \epsilon + \Delta_{21}^{} \left(
	|U_{\tau 2}^{}|^2 - |U_{\mu 2}^{}|^2 \right) +
	\Delta_{31}^{} \left(|U_{\tau 3}^{}|^2 - |U_{\mu 3}^{}|^2
	\right)}{2 \xi} \; .
%              (19)
\end{eqnarray}
This means that the neutrino flavor mixing in dense matter can be approximately described by only one degree of freedom, as having been pointed out in Ref. \cite{Xing:2019owb}:
\begin{eqnarray}
\left. \widetilde{U}\right|_{A\gg\Delta_{31}^{}} \simeq
\begin{pmatrix} 0 & 0 & 1 \cr \cos\theta & ~\sin\theta~ & 0 \cr
-\sin\theta & \cos\theta & 0 \end{pmatrix} \; ,
%     (20)
\end{eqnarray}
where $\theta \in [0, \pi/2]$ and
\begin{eqnarray}
\tan^2\theta=\frac{|\widetilde U_{\mu 2}^{}|^2 }{|\widetilde U_{\mu 1}^{}|^2 }  \simeq  
\frac{  \xi-A \epsilon- \Delta_{21}^{} \left(
	|U_{\tau 2}^{}|^2 - |U_{\mu 2}^{}|^2 \right) -
	\Delta_{31}^{} \left(|U_{\tau 3}^{}|^2 - |U_{\mu 3}^{}|^2
	\right)}{\xi +A \epsilon+ \Delta_{21}^{} \left(
	|U_{\tau 2}^{}|^2 - |U_{\mu 2}^{}|^2 \right) +
	\Delta_{31}^{} \left(|U_{\tau 3}^{}|^2 - |U_{\mu 3}^{}|^2
	\right)} \; .
%              (21)
\end{eqnarray}
Similarly,
the neutrino oscillation probability ${\widetilde P}_{\alpha\beta}^{}$ in Eq. (15) can be approximately written as
\begin{align}
{\widetilde P}_{ee}^{}&\simeq 1,  &{\widetilde P}_{e\mu}^{}&\simeq 0, &{\widetilde P}_{e\tau}^{}&\simeq 0, \nonumber\\
{\widetilde P}_{\mu e}^{} &\simeq 0,  &{\widetilde P}_{\mu\mu}^{} &\simeq 1 - \sin^2 2 \theta \sin^2 \frac{\widetilde\Delta_{21}L}{4 E}, &{\widetilde P}_{\mu\tau}^{}&\simeq \sin^2 2 \theta \sin^2 \frac{\widetilde\Delta_{21}L}{4 E}, \nonumber\\
{\widetilde P}_{\tau e}^{} &\simeq 0,  &{\widetilde P}_{\tau\mu}^{} &\simeq  \sin^2 2 \theta \sin^2 \frac{\widetilde\Delta_{21}L}{4 E}, &{\widetilde P}_{\tau\tau}^{} &\simeq 1 - \sin^2 2 \theta \sin^2 \frac{\widetilde\Delta_{21}L}{4 E}, 
%                 (22)
\end{align}
where $\widetilde \Delta_{21}^{}$ is taken from Eq. (17) and $\sin^2 2\theta=4 |{\widetilde U}_{\mu 1}^{}|^2 (1 -  |{\widetilde U}_{\mu 1}^{}|^2)$ with $|{\widetilde U}_{\mu 1}^{}|^2$ being taken from Eq. (19).
Note that Eq. (22) is similar to Eq. (8) in Ref. \cite{Luo:2019efb} except that we include radiative corrections in $\theta$ and $\widetilde\Delta_{21}^{}$.  In order to numerically test the accuracies,
we define the absolute error of $\widetilde P_{\alpha\beta}^{}$ as 
$\Delta\widetilde P_{\alpha\beta}^{}=| (\widetilde P_{\alpha\beta}^{})_{\rm Exact}^{} - (\widetilde P_{\alpha\beta}^{})_{\rm Approximate}^{}|$, where $(\widetilde P_{\alpha\beta})_{\rm Exact} ^{}$ stand for the exact results of $\widetilde P_{\alpha\beta}^{}$ and $(\widetilde P_{\alpha\beta}^{})_{\rm Approximate}^{}$
represent the approximate results of $\widetilde P_{\alpha\beta}^{}$. The absolute errors of $ \widetilde  P_{\alpha\beta}^{}$ in Eq. (22) with different $L/E$ and $A/\Delta_{31}$ are demonstrated in Fig. 3. Similar to the case without radiative corrections discussed in Ref. \cite{Luo:2019efb}, the analytical expressions of $\widetilde P_{\alpha\beta}^{} $
in Eq. (22) are accurate enough in most of the parameter space. For the upper left part in each
subgraph of Fig. 3, we need to keep higher orders of $\Delta_{21}^{}/A$, $\Delta_{31}^{}/A$ and $\epsilon$, or just make perturbative expansions in terms of
$\Delta_{21}^{}/A$ and $\epsilon$ to improve the accuracies of $\widetilde P_{\alpha\beta}^{} $.

To be more explicit, if the $A\epsilon$ term is not bigger than the $\Delta_{21}^{}$ term in Eq. (19),
we can further simplify $|\widetilde U_{\alpha i}^{}|^2$ (for $\alpha i=\mu 1, \mu2, \tau1,\tau2$) as $|\widetilde{U}_{\mu 1}^{}|^2  \simeq  |\widetilde{U}_{\tau 2}^{}|^2 \simeq |U_{\tau 3}^{}|^2/(|U_{\mu 3}^{}|^2+|U_{\tau 3}^{}|^2)$ and
 $|\widetilde{U}_{\mu 2}^{}|^2  \simeq  |\widetilde{U}_{\tau 1}^{}|^2 \simeq |U_{\mu 3}^{}|^2/(|U_{\mu 3}^{}|^2+|U_{\tau 3}^{}|^2)$. This is equivalent to the asymptotic values of $|\widetilde U_{\alpha i}^{}|^2$ (for $\alpha i=\mu1,\mu2,\tau1,\tau2$) in the $A\to \infty$ limit when radiative corrections are not taken into account (the blue dashed line in Fig. 2).
 As the increase of $A$, the $A\epsilon$ term in Eq. (19) becomes non-negligible. If
the $A\epsilon$ term and $\Delta_{31}^{}$ term are of the same order, the relation
\begin{eqnarray}
\frac{{\rm d }\left(|{\widetilde U}_{\mu 1}|^2\right)}{{\rm d} A}\simeq
-\frac{{\rm d }\left(|{\widetilde U}_{\mu 2}|^2\right)}{{\rm d} A}\simeq\frac{2\left(1 - |U_{\mu 3}^{}|^2\right)|U_{\mu 3}^{}|^2 \epsilon}{\Delta_{31}^{}} \;,
%         (22)
\end{eqnarray}
can be derived.
This means $\theta=\arctan(|\widetilde U_{\mu 2}^{}|/|\widetilde U_{\mu 1}^{}|)$ will decrease with the increase of $A$ due to the existence of the radiative correction parameter $\epsilon$. 
In the $A\to \infty$ limit, it is easy to infer from Eqs. (19) and (21) that $|{\widetilde U}_{\alpha i}^{}|^2$  trivially take $0$ or $1$ and $\theta$ is approaching zero, implying that all the three flavors do not oscillate into one another.  Thus it makes no sense to discuss lepton flavor mixing in this extreme case.
Considering the four cases (NMO, $\nu$), (IMO, $\nu$), (IMO, $\overline\nu$) and (IMO, $\overline\nu$) separately, we summarize
the corresponding analytical expressions of $\widetilde \Delta_{ij}^{}$ (for $ij=21,31$) and $\widetilde U$ in 
the $A\to \infty$ limit in Table 1 while the other three cases will be discussed later. 
%%%%%%%%%%%%%%%%%%%% table 1  %%%%%%%%%%%%%%%%%%%%%%%%
%%%%%%%%%%%%%%%%%%%%%%%%%%%%
\begin{table}[h]
	\caption{In the standard three-flavor mixing scheme, the analytical expressions of $\widetilde \Delta_{ij}^{}$ (for $ij=21,31$) 
		and $\widetilde U$ in the $A\to \infty$ limit, where we only show the terms of  order ${\cal O} (A)$ for $\widetilde \Delta_{ij}^{}$.}
	\vspace{0.1cm}
	\centering
	\renewcommand\arraystretch{1.3}
	\begin{tabular}{c|c|c|c|c} \hline\hline
		&(NMO, $\nu$)&(IMO, $\nu$)&(NMO, $\overline\nu$)&(IMO, $\overline\nu$)\\\hline
		$\widetilde \Delta_{21}^{}$& $A\epsilon$&$A (1-\epsilon)$&$A (1-\epsilon)$&$A\epsilon$\\	\hline
		$\widetilde \Delta_{31}^{}$&$A$& $-A\epsilon$&$A$&$-A(1-\epsilon)$\\\hline
		$\widetilde U $&
		$\begin{pmatrix}
		0&0&1\cr
		1&0&0\cr
		0&1&0\cr
		\end{pmatrix}$& $\begin{pmatrix}
		0&1&0\cr
		0&0&1\cr
		1&0&0 \cr
		\end{pmatrix}$&$\begin{pmatrix}
		1&0&0\cr
		0&0&1\cr
		0&1&0 \cr
		\end{pmatrix}$&$\begin{pmatrix}
		0&0&1\cr
		0&1&0\cr
		1&0&0 \cr
		\end{pmatrix}$\\\hline\hline
	\end{tabular}
\end{table}
%%%%%%%%%%%%%%%%%%%%%%%%%%%%%%%%%%%%%%%%%%%%%%%%%%%%%%%%%%%%%%%%%%%%%%%%%%%%%%%%%%%%%%%
\subsection{ (IMO, $\nu$)}
Given a neutrino beam with inverted mass ordering, the evolution of
$\widetilde \Delta_{ij}^{}$ and $|\widetilde U_{\alpha i}^{}|^2$ with $A$ are illustrated in the lower left panel of Fig. 1 and Fig. 4, respectively. Note that there is no intersections between $\widetilde \Delta_{21}^{}$ and $\widetilde \Delta_{31}^{}$ in cases (IMO, $\nu$) and (IMO, $\overline\nu$) in Fig. 1 with $|\widetilde \Delta_{31}^{}| = -\widetilde \Delta_{31}^{}$ being shown in fact.
In the case (IMO, $\nu$), we also note that $\widetilde\Delta_{21}^{}> |\widetilde\Delta_{31}^{}|$ holds
when the matter effect parameter $A$ is big enough.
Analytically, expanding Eq. (6) in $\Delta_{21}^{}/A$, $\Delta_{31}^{}/A$
and $\epsilon$ directly leads to
\begin{eqnarray}
 \widetilde{\Delta}^{}_{21} &\simeq& A-\frac{1}{2}\left[
A \epsilon+ (1 - 3 |U_{e2}^{}|^2) \Delta_{21}^{}+ (1 - 3 |U_{e3}^{}|^2) \Delta_{31}^{} +\xi
\right] \; ,
\nonumber \\
\widetilde{\Delta}^{}_{31} &\simeq&-\xi 
\; , \hspace{1cm}
%     (24)
\end{eqnarray}
with $\xi$ being defined in Eq. (18). 
Consistent with Fig. 1, $\widetilde\Delta_{31}^{}$ approaches $-A \epsilon$ instead of a constant
value in the $A\to\infty$ limit. To understand the asymptotic behaviors of $|\widetilde U_{\alpha i}^{}|^2$ in the $A\to \infty$ limit shown in Fig. 4, we expand Eq. (10) and get
\begin{eqnarray}
|\widetilde{U}_{e1}^{}|^2  & \simeq&  |\widetilde{U}_{e3}^{}|^2 \simeq
|\widetilde{U}_{\mu 2}^{}|^2 \simeq |\widetilde{U}_{\tau 2}^{}|^2 \simeq 0 \; ,
\quad |\widetilde{U}_{e2}^{}|^2 \simeq 1 \; ,
\nonumber \\
|\widetilde{U}_{\mu 1}^{}|^2  & \simeq & |\widetilde{U}_{\tau 3}^{}|^2 \simeq
\frac{1}{2} - \frac{ A \epsilon+ \Delta_{21}^{} \left(
	|U_{\tau 2}^{}|^2 - |U_{\mu 2}^{}|^2 \right) +
	\Delta_{31}^{} \left(|U_{\tau 3}^{}|^2 - |U_{\mu 3}^{}|^2
	\right)}{2 \xi} \; ,
\nonumber\\
|\widetilde{U}_{\mu 3}^{}|^2  & \simeq&  |\widetilde{U}_{\tau 1}^{}|^2 \simeq
\frac{1}{2} + \frac{A \epsilon + \Delta_{21}^{} \left(
	|U_{\tau 2}^{}|^2 - |U_{\mu 2}^{}|^2 \right) +
	\Delta_{31}^{} \left(|U_{\tau 3}^{}|^2 - |U_{\mu 3}^{}|^2
	\right)}{2 \xi} \; .
%              (25)
\end{eqnarray}
So one can use only one parameter to approximately describe lepton flavor mixing,
\begin{eqnarray}
\left. \widetilde{U}\right|_{A\gg\Delta_{31}} \simeq
\begin{pmatrix} 0 & 1 & 0 \cr \cos\theta & ~0~ & \sin\theta \cr
-\sin\theta & 0 & \cos\theta \end{pmatrix} \; ,
%     (26)
\end{eqnarray}
%%%%%%%%%%%%%%%%%% figure 3 %%%%%%%%%%%%%%%%%%%%%%%
\begin{figure}[H]
	\centering{
		\includegraphics[width=16cm]{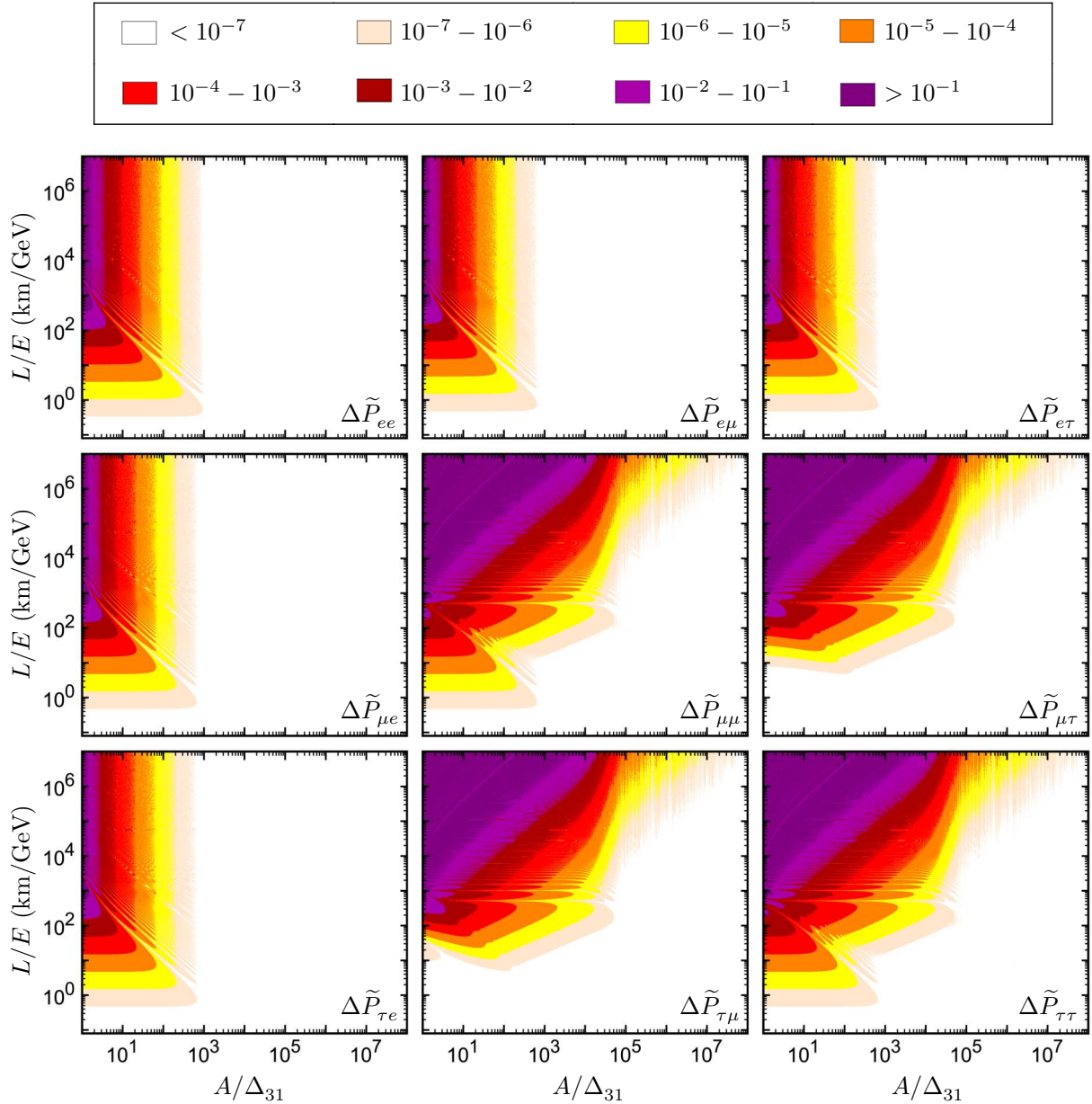}}
	\caption{The absolute errors of our analytical approximations in Eq. (22), where 
		the best-fit values of 
		$(\theta_{12}^{}, \theta_{13}^{}, \theta_{23}^{}, \delta,\Delta_{21}^{}, \Delta_{31}^{} )$ in Ref. \cite{Esteban:2018azc} and $\epsilon\simeq 5 \times 10^{-5}$ \cite{Botella:1986wy} have been input.}
\end{figure}
%%%%%%%%%%%%%%%%%%%%%%%%%%%%%%%%%%%%%%%%%%%%%%%%%%%
%%%%%%%%%%%%%%%%%% figure 4 %%%%%%%%%%%%%%%%%%%%%%%
\begin{figure}[h]
	\centering{
		\includegraphics[width=16cm]{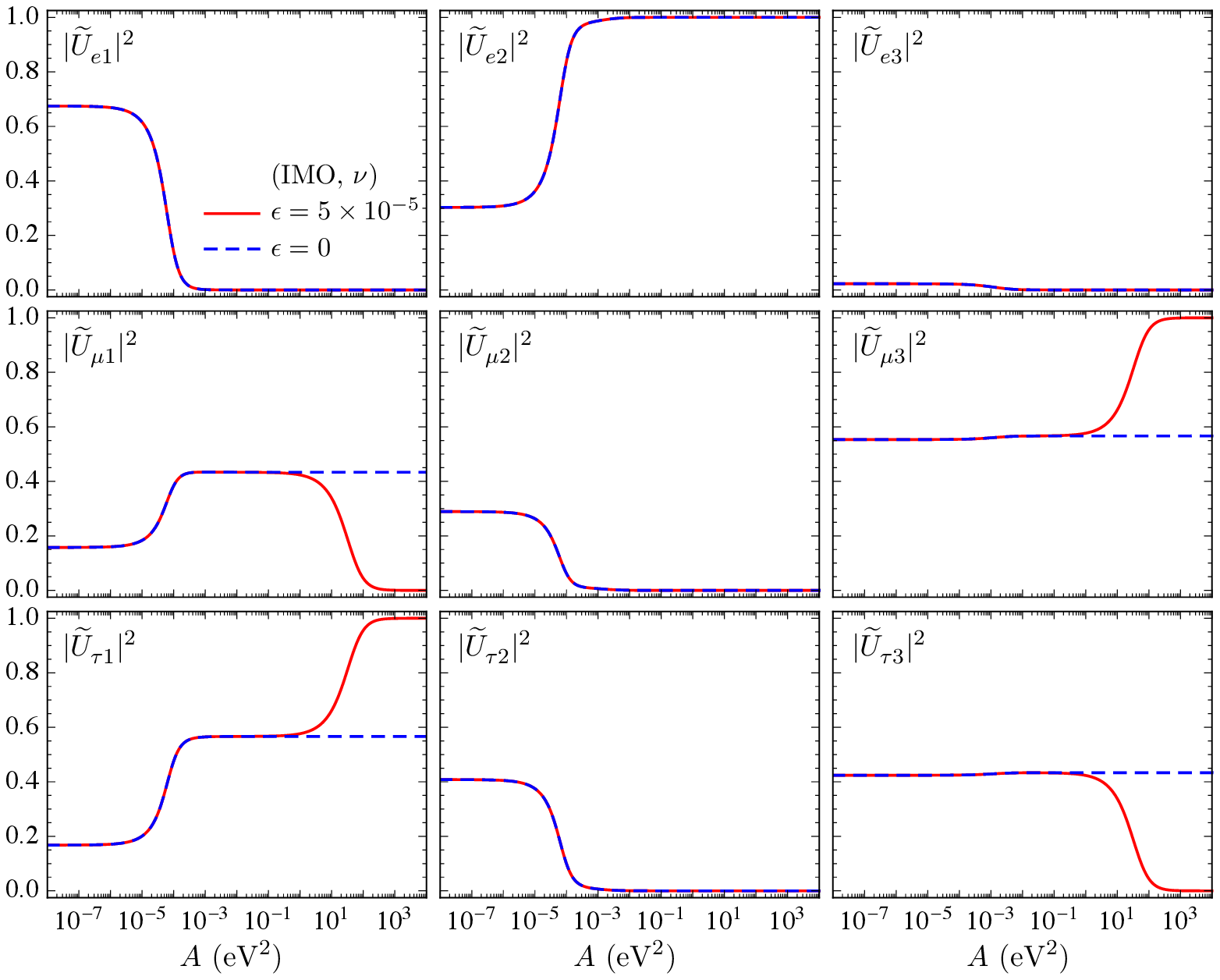}}
	\caption{In the standard three-flavor mixing scheme, the illustration of  how $|\widetilde  U_{\alpha i}^{}|^2$ (for $\alpha=e, \mu, \tau ~{\rm and}~ i=1,2,3$) evolve with the matter effect parameter $A$ in the
		case $({\rm IMO},\nu)$ with or without radiative corrections, where the best-fit values of 
		$(\theta_{12}^{}, \theta_{13}^{}, \theta_{23}^{}, \delta,\Delta_{21}^{}, \Delta_{31}^{} )$ in Ref. \cite{Esteban:2018azc} have been input.}
\end{figure}
%%%%%%%%%%%%%%%%%%%%%%%%%%%%%%%%%%%%%%%%%%%%%%%%%%%
\noindent where $\theta \in[0, \pi/2]$ and
\begin{eqnarray}
\tan^2\theta  = \frac{|\widetilde U_{\mu 3}^{}|^2}{|\widetilde U_{\mu 1}^{}|^2}& \simeq  &
\frac{  \xi + A \epsilon +  \Delta_{21}^{} \left(
	|U_{\tau 2}^{}|^2 - |U_{\mu 2}^{}|^2 \right) +
	\Delta_{31}^{} \left(|U_{\tau 3}^{}|^2 - |U_{\mu 3}^{}|^2
	\right)}{\xi - A \epsilon -  \Delta_{21}^{} \left(
	|U_{\tau 2}^{}|^2 - |U_{\mu 2}^{}|^2 \right) -
	\Delta_{31}^{} \left(|U_{\tau 3}^{}|^2 - |U_{\mu 3}^{}|^2
	\right)} \; .
%              (27)
\end{eqnarray}
The analytical approximations of $\widetilde P_{\alpha\beta}^{}$ in Eq. (15)  turn out to be
\begin{align}
{\widetilde P}_{ee}^{}&\simeq 1,  &{\widetilde P}_{e\mu}^{}&\simeq 0, &{\widetilde P}_{e\tau}^{}&\simeq 0, \nonumber\\
{\widetilde P}_{\mu e}^{} &\simeq 0,  &{\widetilde P}_{\mu\mu}^{} &\simeq 1 - \sin^2 2 \theta \sin^2 \frac{\widetilde\Delta_{31}L}{4 E}, &{\widetilde P}_{\mu\tau}^{}&\simeq \sin^2 2 \theta \sin^2 \frac{\widetilde\Delta_{31}L}{4 E}, \nonumber\\
{\widetilde P}_{\tau e}^{} &\simeq 0,  &{\widetilde P}_{\tau\mu}^{} &\simeq  \sin^2 2 \theta \sin^2 \frac{\widetilde\Delta_{31}L}{4 E}, &{\widetilde P}_{\tau\tau}^{} &\simeq 1 - \sin^2 2 \theta \sin^2 \frac{\widetilde\Delta_{31}L}{4 E}, 
%                 (28)
\end{align}
where $\widetilde \Delta_{31}^{}$ comes from Eq. (24) and $\sin^2 2\theta =4 |\widetilde U_{\mu 1}^{}|^2 (1 - |\widetilde U_{\mu 1}^{}|^2)$ with $|\widetilde U_{\mu 1}^{}|^2$ coming from Eq. (25). For simplicity, we do not show the accuracies of Eq. (28), which are very similar to the case (NMO, $\nu$)  in Fig. 3. 
Comparing Eq. (28) with Eq. (22), we find that it is impossible to discriminate the normal
mass ordering from the inverted mass ordering from neutrino oscillations if the matter density is very big.

We also notice that if the $\Delta_{21}$ term and $A\epsilon$ term in Eq. (25) are of the same order,  one can omit them
and get $|\widetilde{U}_{\mu 1}^{}|^2  \simeq  |\widetilde{U}_{\tau 3}^{}|^2 \simeq |U_{\tau 3}^{}|^2/(|U_{\mu 3}^{}|^2+|U_{\tau 3}^{}|^2)$ and
$|\widetilde{U}_{\mu 3}^{}|^2  \simeq  |\widetilde{U}_{\tau 1}^{}|^2 \simeq |U_{\mu 3}^{}|^2/(|U_{\mu 3}^{}|^2+|U_{\tau 3}^{}|^2)$. This corresponds to the fixed values of $|\widetilde U_{\alpha i}^{}|^2$ (for $\alpha i=\mu1,\mu3,\tau1,\tau3$) in the $A\to \infty$ limit if radiative corrections are not taken into account (the blue dashed line in Fig. 4).
As the increase of $A$, the $A\epsilon$ term will gradually dominate and the neutrino oscillation
behaviors can be very sensitive to $A$.
In the limit of $A \to \infty$, $\theta$ approaches $\pi/2$ and there will be no
neutrino oscillation phenomenon.
\subsection{ (NMO, $\overline\nu$)}
Considering an antineutrino beam with normal mass ordering, we
make the replacements  $A\to-A$ and $U\to-U^{*}$ in Eqs. (5) and (10), and
draw the corresponding evolution of $\widetilde \Delta_{ij}^{}$ and $|\widetilde U_{\alpha i}^{}|^2$ 
with A in the upper right panel of Fig. 1 and Fig. 5, respectively. Note that we always have $\widetilde\Delta_{31}^{}>\widetilde 
\Delta_{21}^{}$ in this case although the difference between them is too small to be shown clearly in
Fig. 1 if $A$ is big enough. The radiative corrections to both $\widetilde\Delta_{21}^{}$ and $\widetilde\Delta_{31}^{}$
are very small, which can be analytically understood. By performing perturbative expansions, $\widetilde\Delta_{ij}^{}$
(for $ij=21,31$) in Eq. (5) are reduced to
%%%%%%%%%%%%%%%%%% figure 5 %%%%%%%%%%%%%%%%%%%%%%%
\begin{figure}[h]
	\centering{
		\includegraphics[width=16cm]{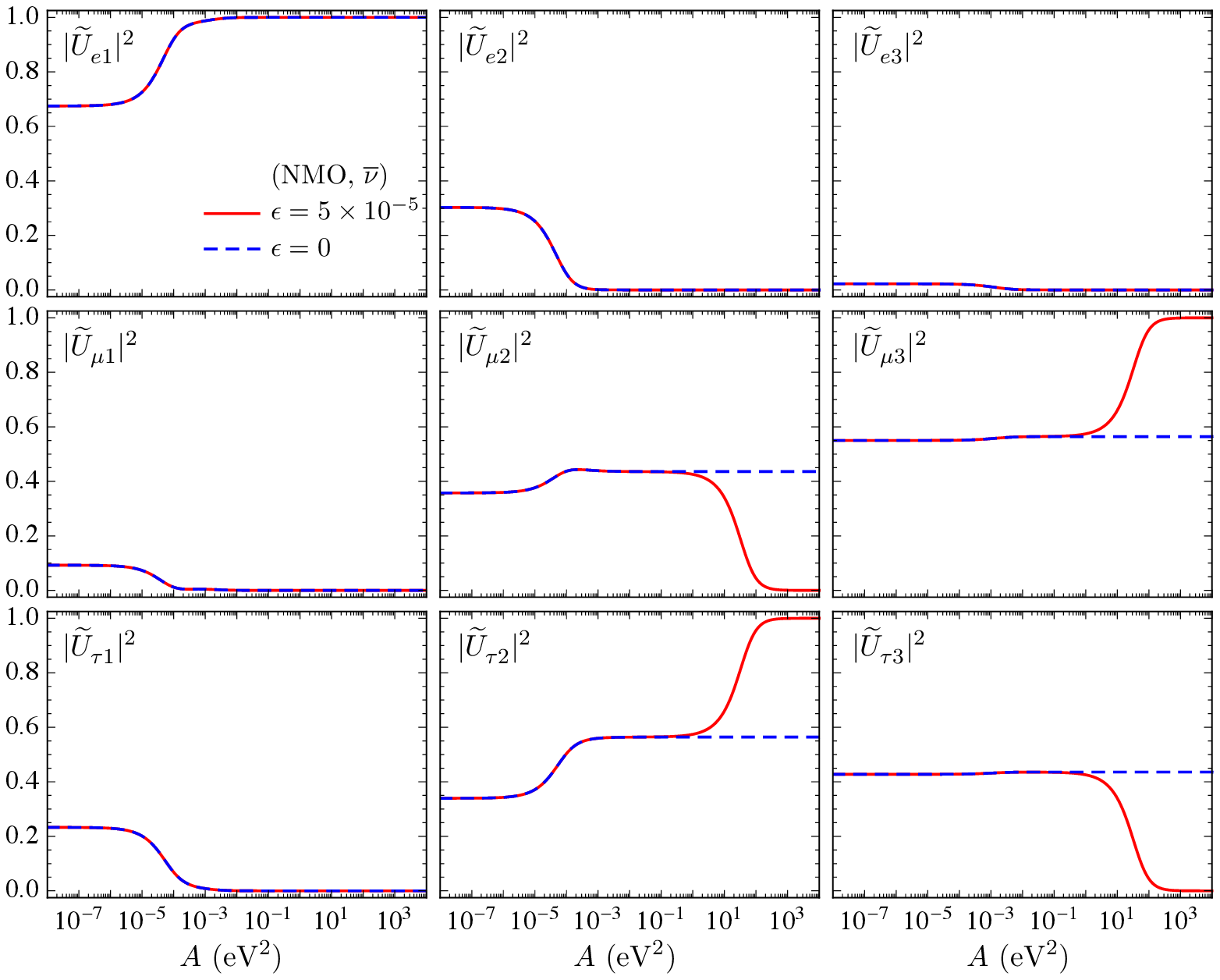}}
	\caption{In the standard three-flavor mixing scheme, the illustration of  how $|\widetilde  U_{\alpha i}^{}|^2$ (for $\alpha=e, \mu, \tau~{\rm and}~ i=1,2,3$) evolve with the matter effect parameter $A$ in the
		case $({\rm NMO},\overline\nu)$ with or without radiative corrections, where the best-fit values of 
		$(\theta_{12}^{}, \theta_{13}^{}, \theta_{23}^{}, \delta,\Delta_{21}^{}, \Delta_{31}^{} )$ in Ref. \cite{Esteban:2018azc} have been input.}
\end{figure}
%%%%%%%%%%%%%%%%%%%%%%%%%%%%%%%%%%%%%%%%%%%%%%%%%%%
\begin{eqnarray}
 \widetilde{\Delta}^{}_{21} &\simeq& A-\frac{1}{2}\left[
A \epsilon- (1 - 3 |U_{e2}^{}|^2) \Delta_{21}^{}- (1 - 3 |U_{e3}^{}|^2) \Delta_{31}^{} +\xi
\right] \; ,
\nonumber \\
\widetilde{\Delta}^{}_{31} &\simeq& A-\frac{1}{2}\left[
A \epsilon- (1 - 3 |U_{e2}^{}|^2) \Delta_{21}^{}- (1 - 3 |U_{e3}^{}|^2) \Delta_{31}^{} -\xi
\right]
%     (29)
\end{eqnarray}
with
\begin{eqnarray}
\xi&=&\big\{
\left[\left( 1 - |U_{e2}^{}|^2  \right)\Delta_{21}^{}
+\left( 1 - |U_{e3}^{}|^2 \right) \Delta_{31}^{} - A\epsilon\right]^2
+ 4 A \epsilon\left( |U_{\mu 2}^{}|^2 \Delta_{21} + |U_{\mu 3}^{}|^2 \Delta_{31} \right) \nonumber\\
&&- 4 |U_{e1}^{}|^2 \Delta_{21}^{} \Delta_{31}^{}
\big\}^{1/2}\;.
%              (30)
\end{eqnarray}
From Eq. (29), it is clear that
the leading order of $\widetilde\Delta_{21}^{}$ and $\widetilde\Delta_{31}^{}$ is $A$ and the
radiative corrections in the next-to-leading order do not matter a lot.
The only difference between Eq. (30) (the expression of $\xi$ in the case (NMO, $\overline\nu$)) and Eq. (18)  (the expression of $\xi$ in the case (NMO, $\nu$)) is the sign of $\epsilon$.  Similarly, $|\widetilde U_{\alpha i}^{}|^2$ can be
expanded as
\begin{eqnarray}
|\widetilde{U}_{e2}^{}|^2  & \simeq & |\widetilde{U}_{e3}^{}|^2 \simeq
|\widetilde{U}_{\mu 1}^{}|^2 \simeq |\widetilde{U}_{\tau 1}^{}|^2 \simeq 0 \; ,
\quad |\widetilde{U}_{e1}^{}|^2 \simeq 1 \; ,
\nonumber \\
|\widetilde{U}_{\mu 2}^{}|^2  & \simeq & |\widetilde{U}_{\tau 3}^{}|^2 \simeq
\frac{1}{2} - \frac{ A \epsilon-\Delta_{21}^{} \left(
	|U_{\tau 2}^{}|^2 - |U_{\mu 2}^{}|^2 \right) -
	\Delta_{31}^{} \left(|U_{\tau 3}^{}|^2 - |U_{\mu 3}^{}|^2
	\right)}{2 \xi} \; ,
\nonumber\\
|\widetilde{U}_{\mu 3}^{}|^2  & \simeq  &|\widetilde{U}_{\tau 2}^{}|^2 \simeq
\frac{1}{2} + \frac{A \epsilon - \Delta_{21}^{} \left(
	|U_{\tau 2}^{}|^2 - |U_{\mu 2}^{}|^2 \right) -
	\Delta_{31}^{} \left(|U_{\tau 3}^{}|^2 - |U_{\mu 3}^{}|^2
	\right)}{2 \xi} \; ,
%              (31)
\end{eqnarray}
namely,
\begin{eqnarray}
\left. \widetilde{U}\right|_{A\gg\Delta_{31}^{}} \simeq
\begin{pmatrix} 1 & 0 & 0 \cr 0 & \cos\theta & \sin\theta \cr
0 & -\sin\theta & \cos\theta \end{pmatrix} \; ,
%     (32)
\end{eqnarray}
where $\theta \in [0,\pi/2]$ and
\begin{eqnarray}
\tan^2\theta =\frac{|\widetilde U_{\mu 3}^{}|^2}{|\widetilde U_{\mu 2}^{}|^2} & \simeq & 
\frac{ \xi + A \epsilon -  \Delta_{21}^{} \left(
	|U_{\tau 2}^{}|^2 - |U_{\mu 2}^{}|^2 \right) -
	\Delta_{31}^{} \left(|U_{\tau 3}^{}|^2 - |U_{\mu 3}^{}|^2
	\right)}{\xi - A \epsilon +  \Delta_{21}^{} \left(
	|U_{\tau 2}^{}|^2 - |U_{\mu 2}^{}|^2 \right) +
	\Delta_{31}^{} \left(|U_{\tau 3}^{}|^2 - |U_{\mu 3}^{}|^2
	\right)} \; .
%              (33)
\end{eqnarray}
The corresponding neutrino oscillation probabilities in matter are approximately
\begin{align}
{\widetilde P}_{ee}^{}&\simeq 1,  &{\widetilde P}_{e\mu}^{}&\simeq 0, &{\widetilde P}_{e\tau}^{}&\simeq 0, \nonumber\\
{\widetilde P}_{\mu e}^{} &\simeq 0,  &{\widetilde P}_{\mu\mu}^{} &\simeq 1 - \sin^2 2 \theta \sin^2 \frac{\widetilde\Delta_{32}L}{4 E}, &{\widetilde P}_{\mu\tau}^{}&\simeq \sin^2 2 \theta \sin^2 \frac{\widetilde\Delta_{32}L}{4 E}, \nonumber\\
{\widetilde P}_{\tau e}^{} &\simeq 0,  &{\widetilde P}_{\tau\mu}^{} &\simeq  \sin^2 2 \theta \sin^2 \frac{\widetilde\Delta_{32}L}{4 E}, &{\widetilde P}_{\tau\tau}^{} &\simeq 1 - \sin^2 2 \theta \sin^2 \frac{\widetilde\Delta_{32}L}{4 E}, 
%                 (30)
\end{align}
where $\sin^2 2 \theta=|\widetilde U_{\mu 2}^{}|^2 (1 - |\widetilde U_{\mu 2}^{}|^2)$ and 
$\widetilde \Delta_{32}^{} \simeq \xi$ from Eq. (29). The accuracies of ${\widetilde P}_{\alpha\beta}^{}$ in
Eq. (34)
are similar to the case (NMO, $\nu$) in Fig. 3. 

If $A$ is small enough, we can ignore the smaller terms of $\Delta_{21}$ and $A\epsilon$ in Eq. (31),
and obtain $|\widetilde{U}_{\mu 2}^{}|^2  \simeq  |\widetilde{U}_{\tau 3}^{}|^2 \simeq |U_{\tau 3}^{}|^2/(|U_{\mu 3}^{}|^2+|U_{\tau 3}^{}|^2)$ and
$|\widetilde{U}_{\mu 3}^{}|^2  \simeq  |\widetilde{U}_{\tau 2}^{}|^2 \simeq |U_{\mu 3}^{}|^2/(|U_{\mu 3}^{}|^2+|U_{\tau 3}^{}|^2)$. This is consistent with the fixed values of $|\widetilde U_{\alpha i}^{}|^2$ (for $\alpha i=\mu2,\mu3,\tau2,\tau3$) in the $A\to \infty$ limit if radiative corrections are not included (the blue dashed line in Fig. 5).
If the term $A\epsilon$ too big to be abandoned,  the neutrino flavor mixing can be significantly affected by $A$.
In the $A\to \infty$ limit, $|\widetilde U_{\alpha i}^{}|^2$
trivially take 0 or 1 and $\theta$ approaches $\pi/2$, leading to no neutrino oscillations.
\subsection{(IMO, $\overline\nu$)}
Similarly, in the case (IMO, $\overline\nu$), i.e. an antineutrino beam with inverted mass
ordering, we illustrate the evolution of $\widetilde \Delta_{ij}^{}$ and $|\widetilde U_{\alpha i}^{}|^2$ in
the lower right panel of Fig. 1 and Fig. 6, respectively. Through making perturbative expansions, 
$\widetilde \Delta_{ij}^{}$ and $|\widetilde U_{\alpha i}^{}|^2$ can be approximately expressed as
%%%%%%%%%%%%%%%%%% figure 6 %%%%%%%%%%%%%%%%%%%%%%%
\begin{figure}[h]
	\centering{
		\includegraphics[width=16cm]{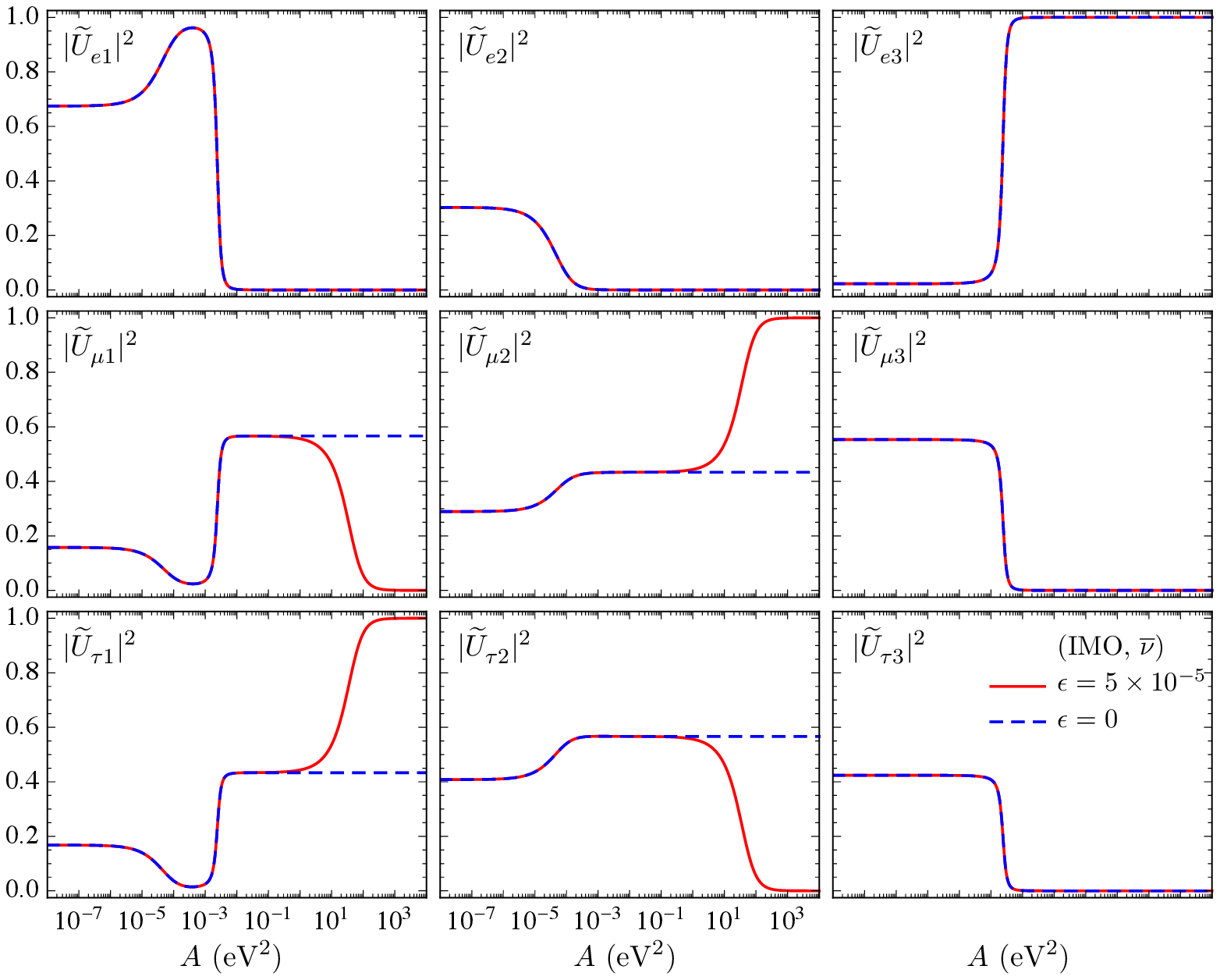}}
	\caption{In the standard three-flavor mixing scheme, the illustration of  how $|\widetilde  U_{\alpha i}^{}|^2$ (for $\alpha=e, \mu, \tau~{\rm and} ~ i=1,2,3$) evolve with the matter effect parameter $A$ in the
		case $({\rm IMO},\overline\nu)$ with or without radiative corrections, where the best-fit values of 
		$(\theta_{12}^{}, \theta_{13}^{}, \theta_{23}^{}, \delta,\Delta_{21}^{}, \Delta_{31}^{} )$ in Ref. \cite{Esteban:2018azc} have been input.}
\end{figure}
%%%%%%%%%%%%%%%%%%%%%%%%%%%%%%%%%%%%%%%%%%%%%%%%%%%
\begin{eqnarray}
 \widetilde{\Delta}^{}_{21} &\simeq& \xi
\; ,
\nonumber \\
\widetilde{\Delta}^{}_{31} &\simeq& -A+\frac{1}{2}\left[
A \epsilon- (1 - 3 |U_{e2}^{}|^2) \Delta_{21}^{}- (1 - 3 |U_{e3}^{}|^2) \Delta_{31}^{} +\xi
\right]
\; , \hspace{1cm}
%     (35)
\end{eqnarray}
and
\begin{eqnarray}
|\widetilde{U}_{e1}^{}|^2  & \simeq&  |\widetilde{U}_{e2}^{}|^2 \simeq
|\widetilde{U}_{\mu 3}^{}|^2 \simeq |\widetilde{U}_{\tau 3}^{}|^2 \simeq 0 \; ,
\quad |\widetilde{U}_{e3}^{}|^2 \simeq 1 \; ,
\nonumber \\
|\widetilde{U}_{\mu 1}^{}|^2  & \simeq & |\widetilde{U}_{\tau 2}^{}|^2 \simeq
\frac{1}{2} - \frac{ A \epsilon-\Delta_{21}^{} \left(
	|U_{\tau 2}^{}|^2 - |U_{\mu 2}^{}|^2 \right) -
	\Delta_{31}^{} \left(|U_{\tau 3}^{}|^2 - |U_{\mu 3}^{}|^2
	\right)}{2 \xi} \; ,
\nonumber\\
|\widetilde{U}_{\mu 2}^{}|^2  & \simeq&  |\widetilde{U}_{\tau 1}^{}|^2 \simeq
\frac{1}{2} + \frac{A \epsilon - \Delta_{21}^{} \left(
	|U_{\tau 2}^{}|^2 - |U_{\mu 2}^{}|^2 \right) -
	\Delta_{31}^{} \left(|U_{\tau 3}^{}|^2 - |U_{\mu 3}^{}|^2
	\right)}{2 \xi} \; ,
%              (36)
\end{eqnarray}
where $\xi$ has been defined in Eq. (30). From either Fig. 1 or Eq. (35),  it is clear that the radiative corrections to $\widetilde\Delta_{21}$ 
are  very important  in this case if the matter density is big enough. According to Eq. (36), the lepton flavor mixing matrix
can be approximately parametrized as:
 \begin{eqnarray}
\left. \widetilde{U}\right|_{A \gg \Delta_{31}^{}} \simeq
\begin{pmatrix} 0 & 0 & ~1 \cr \cos\theta & \sin\theta & ~0 \cr
-\sin\theta & \cos\theta & ~0 \end{pmatrix} \; ,
%     (37)
\end{eqnarray}
where $\theta\in[0, \pi/2]$ and
\begin{eqnarray}
\tan^2\theta  & \simeq  &
\frac{  \xi + A \epsilon - \Delta_{21}^{} \left(
	|U_{\tau 2}^{}|^2 - |U_{\mu 2}^{}|^2 \right) -
	\Delta_{31}^{} \left(|U_{\tau 3}^{}|^2 - |U_{\mu 3}^{}|^2
	\right)}{\xi - A \epsilon + \Delta_{21}^{} \left(
	|U_{\tau 2}^{}|^2 - |U_{\mu 2}^{}|^2 \right) +
	\Delta_{31}^{} \left(|U_{\tau 3}^{}|^2 - |U_{\mu 3}^{}|^2
	\right)} \; .
%              (38)
\end{eqnarray}
The corresponding analytical approximations of $\widetilde P_{\alpha\beta}^{}$ are the same as
Eq. (22) except that $\widetilde \Delta_{21}^{}$ and $\sin^2 2\theta=
|\widetilde U_{\mu1}^{}|^2 (1 - |\widetilde U_{\mu1}^{}|^2)$ should be taken from Eq. (35) and Eq. (36),
respectively.
Comparing Eq. (22) and Eq. (34), we find that it is impossible to discriminate between the normal
mass ordering and the inverted mass ordering from antineutrino oscillations if the matter density is very big.

If the $\Delta_{21}$ term and the $A\epsilon$ term in Eq. (36) are of the same order, we can omit them and arrive at $|\widetilde{U}_{\mu 1}^{}|^2  \simeq  |\widetilde{U}_{\tau 2}^{}|^2 \simeq |U_{\mu 3}^{}|^2/(|U_{\mu 3}^{}|^2+|U_{\tau 3}^{}|^2)$ and
$|\widetilde{U}_{\mu 2}^{}|^2  \simeq  |\widetilde{U}_{\tau 1}^{}|^2 \simeq |U_{\tau 3}^{}|^2/(|U_{\mu 3}^{}|^2+|U_{\tau 3}^{}|^2)$. This coincides with the fixed values of $|\widetilde U_{\alpha i}^{}|^2$ (for $\alpha i=\mu1,\mu2,\tau1,\tau2$) in the $A\to \infty$ limit if radiative corrections are not included (the blue dashed line in Fig. 6).
If the term $A\epsilon$ is too big to be omitted,  it will affect the neutrino flavor mixing a lot.
 In the $A \to \infty $, $\theta$ approaches $\pi/2$ and no neutrino oscillations between
$\nu_e^{}$, $\nu_\mu^{}$ and $\nu_\tau^{}$ will happen.
%%%%%%%%%%%%%%%%%%%    section     3    %%%%%%%%%%%%%%%%%%%%%%
%%%%%%%%%%%%%%%%%%%%%%%%%%%%%%%%%%%%%%%%%%%%%%%%%%%%
\section{The $(3+1)$ flavor mixing scheme}
Now we turn to the $(3+1)$ flavor mixing scheme with one more light sterile neutrino.
The corresponding Hamiltonian ${\cal H}_{\rm m}^s$ describing the propagation of neutrinos in matter has 
been shown in Eq. (4). Analogous to Ref. \cite{Li:2018ezt}, the eigenvalues $\lambda$ of ${\cal H}_{\rm m}^s$
can be derived by solving the equation
\begin{eqnarray}
\lambda^4 + b \lambda^3 + c \lambda^2 + d \lambda +e=0 \;,
%              (39)
\end{eqnarray}
where the expressions of the relevant coefficients are as follows 
\begin{eqnarray}
b&=& -\sum_i^{} \Delta_{i1}^{} - A (1 + \epsilon) -A^{\prime} \;,
\nonumber \\
c &=& \sum_{i<j}^{} \Delta_{i1}^{} \Delta_{j1} 
+ A \sum_{i}^{} \Delta_{i1} (1 - |V_{e i}^{}|^2)
+ \epsilon A \sum_{i}^{} \Delta_{i1} (1 - |V_{\tau  i}^{}|^2)
\nonumber\\
&&+ A^{\prime} \sum_{i}^{} \Delta_{i1} (1 - |V_{s i}^{}|^2)
+ A A^{\prime} (1 + \epsilon) + A^2 \epsilon \;,
\nonumber\\
d &=& -\Delta_{21}^{}\Delta_{31}^{}\Delta_{41}^{}
-\sum_{i,j,k}^{} \frac{\varepsilon^2_{ijk}}{2} \Delta_{i1}^{}
\Delta_{j1}^{} (A |V_{e k}^{}|^2 + \epsilon A |V_{\tau k}^{}|^2 
+ A^{\prime} |V_{sk}^2|) \nonumber\\
&& -A A^{\prime} \sum_{i}^{} \Delta_{i1}^{} (|V_{\mu i}^{}|^2 +|V_{\tau i}^{}|^2)-\epsilon A^{2} \sum_{i}^{} \Delta_{i1}^{} (|V_{\mu i}^{}|^2 +|V_{s i}^{}|^2)  \nonumber\\
&&- \epsilon A A^{\prime} \sum_{i}^{} \Delta_{i1}^{} (|V_{e  i}^{}|^2 +|V_{\mu i}^{}|^2) - \epsilon A^2 A^{\prime} \;,\nonumber\\
e &=& \Delta_{21}^{}\Delta_{31}^{}\Delta_{41}^{} (A |V_{e1}^{}|^2 + \epsilon
A |V_{\tau 1}^{}|^2 + A^{\prime} |V_{s1}^{}|^2) + AA^{\prime} \sum_{i,j,k,l}^{} \frac{\varepsilon^2_{ijkl}}{4} \Delta_{i1}^{} \Delta_{j1}^{}
C_{es}^{kl} \nonumber \\
&& + \epsilon A^{2} \sum_{i,j,k,l}^{} \frac{\varepsilon^2_{ijkl}}{4} \Delta_{i1}^{} \Delta_{j1}^{}
C_{e\tau}^{kl}  
+ \epsilon AA^{\prime} \sum_{i,j,k,l}^{} \frac{\varepsilon^2_{ijkl}}{4} \Delta_{i1}^{} \Delta_{j1}^{}
C_{s\tau}^{kl}    
+ \epsilon A^2 A^{\prime} \sum_{i}^{} \Delta_{i1}^{} |V_{\mu i}^{}|^2
%C_i^{} =|V_{\mu i}^{}|^2
%                   (40)
\end{eqnarray}
with $i,j,k,l=1,2,3,4$; $\varepsilon_{ijkl}^{}$ being four-dimension Levi-Civita symbol; and
\begin{eqnarray}
C_{\alpha\beta}^{ij} =
|V_{\alpha i}^{} V_{\beta j}^{} - V_{\alpha j}^{} V_{\beta i}^{}|^2
%\;, \quad C_i^{}=|\sum_{j,k,l}^{} \varepsilon_{ijkl}^{}V_{ej}^{} V_{\tau k}^{} V_{sl}^{}|^2
%                  (41)
\end{eqnarray}
for $\alpha\beta=es, e\tau, s\tau$. By defining $\lambda_1^{}<\lambda_2^{}<\lambda_3^{}<\lambda_4^{}$ for the normal
mass ordering and $\lambda_3^{}<\lambda_1^{}<\lambda_2^{}<\lambda_4^{}$
for the inverted ordering, we have $\widetilde \Delta_{ij}^{} =\lambda_i^{} - \lambda_j^{}$. 
After a tedious but straightforward calculation, the exact expressions of $\widetilde \Delta_{ij}^{}$ (for $ij=21,31,41$)
are \cite{Li:2018ezt}
%and defining $\lambda_1^{}<\lambda_2^{}<\lambda_3^{}<\lambda_4^{}$
\begin{eqnarray}
\widetilde \Delta_{21}^{} &=& \sqrt{-4 S^2 -2p+\frac{q}{S}} \;, \nonumber\\
\widetilde \Delta_{31}^{} &=& 2S + \frac{1}{2} \left(\sqrt{-4 S^2 -2p+\frac{q}{S}}-\sqrt{-4 S^2 -2p -\frac{q}{S}}\right)\;, \nonumber\\
\widetilde \Delta_{41}^{} &=& 2S + \frac{1}{2} \left(\sqrt{-4 S^2 -2p+\frac{q}{S}} + \sqrt{-4 S^2 -2p -\frac{q}{S}}\right) 
\end{eqnarray}
for the normal mass ordering ($ m_1 < m_2 <m_3 <m_4 $) and
\begin{eqnarray}
\widetilde \Delta_{21}^{} &=& 2S - \frac{1}{2} \left(\sqrt{-4 S^2 -2p+\frac{q}{S}} + \sqrt{-4 S^2 -2p -\frac{q}{S}}\right)  \;, \nonumber\\
\widetilde \Delta_{31}^{} &=& -\sqrt{-4 S^2 -2p+\frac{q}{S}}\;, \nonumber\\
\widetilde \Delta_{41}^{} &=& 2S - \frac{1}{2} \left(\sqrt{-4 S^2 -2p+\frac{q}{S}}-\sqrt{-4 S^2 -2p -\frac{q}{S}}\right) 
%              (43)
\end{eqnarray}
for the inverted mass ordering ($ m_3 < m_1 <m_2 <m_4 $),
where
\begin{eqnarray}
p &=&c-\frac{ 3 b^2}{8}\;, \quad q=\frac{b^3}{8}-\frac{1}{2} b c + d \;,\nonumber\\
S&= &\frac{1}{2} \sqrt{-\frac{2}{3} p +\frac{2}{3} \sqrt{c^2-3bd + 12 e } \cos \frac{1}{3}\left[ \arccos \left(\frac{2 c^3 -9 bcd +27 b^2 e +27 d^2 -72 ce}{ 2 (c^2-3bd + 12 e)^{3/2}}\right)\right]} \;.
\nonumber\\
%       (44)
\end{eqnarray}
Note that the formulas of $\widetilde\Delta_{ij}^{}$ in Eq. (42) are the same as Eq. (3.4) in Ref. \cite{Li:2018ezt} except that
the coefficients $b,c,d$ and $e$ in Eq. (40) include radiative corrections.
By taking the trace of ${\cal H}_{\rm m}^s$, $B$ can be expressed as
\begin{eqnarray}
B= \frac{1}{4} \left[\Delta_{21}^{} + \Delta_{31}^{} + \Delta_{41}^{}  + A (1+\epsilon) +A^{\prime}-\widetilde \Delta_{21}^{} -  \widetilde\Delta_{31}^{} - \widetilde \Delta_{41}^{}\right].
\end{eqnarray}
By considering the unitarity conditions of $\widetilde V$ and the sum rules derived from ${\cal H}_{\rm m}^s$, $({\cal H}_{\rm m}^s)^2$ and
$({\cal H}_{\rm m}^s)^3$, we get a full set of linear equations of $\widetilde V_{\alpha i}^{}\widetilde V_{\beta i}^{*}$ (for $i=1,2,3,4$ and $\alpha,\beta=e,\mu,\tau,s$),
\begin{eqnarray}
&&\sum_i^{} \widetilde V_{\alpha i}^{} \widetilde V_{\beta i}^{*}= \delta_{\alpha\beta} \; ,
\nonumber \\
&&\sum_i^{} \widetilde \Delta_{i1}^{}\widetilde V_{\alpha i}^{} \widetilde V_{\beta i}^{*}= 
\sum_i^{}  \Delta_{i1}^{}V_{\alpha i}^{} V_{\beta i}^{*} + {\cal A}_{\alpha\beta}^{} - B \delta_{\alpha\beta}
\;, \nonumber \\
&&\sum_i^{} \widetilde \Delta_{i1}^{}\left(\widetilde \Delta_{i1}^{} + 2 B\right)\widetilde V_{\alpha i}^{} \widetilde V_{\beta i}^{*}= 
\sum_i^{}  \Delta_{i1}^{}\left(\Delta_{i1}^{} + {\cal A}_{\alpha\alpha} + {\cal A}_{\beta\beta}\right)V_{\alpha i}^{} V_{\beta i}^{*}  + {\cal A}_{\alpha\beta}^2- B^2 \delta_{\alpha\beta} \;, \nonumber\\
&& \sum_i^{}  \widetilde \Delta_{i1}^{}\left(\widetilde \Delta_{i1}^{2} + 3 B \widetilde \Delta_{i1}^{} + 3 B^2\right)\widetilde V_{\alpha i}^{} \widetilde V_{\beta i}^{*} =  \sum_i^{}
\bigg[\Delta_{i1}^3 + \frac{3}{2} \Delta_{i1}^2 \left({\cal A}_{\alpha\alpha}^{} + {\cal A}_{\beta\beta}^{}\right) + \Delta_{i1}^{} \left({\cal A}_{\alpha\alpha}^{2} 
\right.\nonumber\\&&\left. \hspace{7.15cm} + {\cal A}_{\beta\beta}^{2}
 +{ \cal A}_{\alpha\alpha}^{}  {\cal A}_{\beta\beta}^{}\right) +{\cal A}_{\alpha\beta}^3\bigg]V_{\alpha i}^{} V_{\beta i}^{*} \nonumber\\&& \hspace{7.15cm} 
-B^3 \delta_{\alpha\beta}^{} + C_{\alpha\beta}^{} \;,
%      (46)
\end{eqnarray}
where ${\cal A}_{\alpha\beta}^{} $ denotes the $(\alpha,\beta)$ element of the matter potential matrix ${\cal A}={\rm Diag} \{A, 0, A\epsilon, A^{\prime}\}$ and
\begin{eqnarray}
C_{\alpha\beta}^{} &=& -\frac{1}{2} \sum_{m,n,\gamma}^{} \Delta_{mn}^2 V_{\alpha m}^{} V_{\gamma m}^{*} V_{\gamma n}^{} V_{\beta n}^{*} {\cal A}_{\gamma\gamma}^{}
%               (47)
\end{eqnarray}
with $m,n=1,2,3,4$ and $\alpha,\beta,\gamma=e,\mu,\tau,s$.
According to Eq. (46), one can directly derive the exact expressions of $|\widetilde V_{\alpha i}^{}|^2$
and $\widetilde V_{\alpha i}^{} \widetilde V_{\beta i}^{*}$, which have been given in Refs. \cite{Li:2018ezt, Zhang:2006yq}. We rewrite them as
\begin{eqnarray}
|\widetilde V_{\alpha i}^{}|^2 &=& \frac{1}{\prod \limits^{}_{k\neq i}\widetilde \Delta_{ik}^{}} \sum_j^{} \left(F_{\alpha}^{ij}  |V_{\alpha j}^{}|^2 +C_{\alpha\alpha} \right) \;
%                 (48)
\end{eqnarray}
with
\begin{eqnarray}
F_{\alpha}^{ij} &=& \prod \limits^{}_{k\neq i} \left( \Delta_{j1}^{} - \widetilde\Delta_{k1}^{} - B+{\cal A}_{\alpha\alpha}^{}\right)\nonumber\\ \;.
%                 (49)
\end{eqnarray}
and
\begin{eqnarray}
\widetilde V_{\alpha i}^{} \widetilde V_{\beta i}^{*}&=& \frac{1}{\prod \limits^{}_{k\neq i}\widetilde \Delta_{ik}^{}} \sum_j^{} \left(F_{\alpha\beta}^{ij}  V_{\alpha j}^{} V_{\beta j}^{*}+C_{\alpha\beta}^{}\right)
%                (50)
\end{eqnarray}
with
\begin{eqnarray}
F_{\alpha\beta}^{ij} &=& \frac{1}{2}\bigg[- \left( \Delta_{j1}^{} - \widetilde\Delta_{i1}^{} - B\right) \left({\cal A}_{\alpha\alpha}^2 +{\cal A}_{\beta\beta}^{2} + 4 {\cal A}_{\alpha\alpha}^{} {\cal A}_{\beta\beta}^{}\right)\nonumber\\&& +\prod \limits^{}_{k\neq i} \left( \Delta_{j1}^{} - \widetilde\Delta_{k1}^{} - B + {\cal A}_{\alpha\alpha}^{} + {\cal A}_{\beta\beta}^{}\right) + \prod \limits^{}_{k\neq i} \left( \Delta_{j1}^{} - \widetilde\Delta_{k1}^{} - B\right) \bigg]\;,
%           (51)
\end{eqnarray}
where $\alpha,\beta=e,\mu,\tau,s$; $i,j,k=1,2,3,4$, $B$ is taken from Eq. (45). Comparing with the formulas of
$|\widetilde V_{\alpha i}^{}|^2$
and $\widetilde V_{\alpha i}^{} \widetilde V_{\beta i}^{*}$ in Refs. \cite{Li:2018ezt, Zhang:2006yq}, we get rid of the uneasy terms  $m_i^2 - \widetilde m_j^2$ and include the radiative correction effects in Eqs. (48) and (50).
With the help of Eqs. (42), (43) and (50), we can directly write out the probability of $\nu_{\alpha}^{} \to \nu_{\beta}^{}$ (for $\alpha,\beta=e,\mu,\tau,s$) in a medium:
\begin{eqnarray}
\widetilde P_{\alpha\beta}^{} &=& \delta_{\alpha\beta} - \sum_{i<j}^{} {\rm Re} \left( \widetilde V_{\alpha i}^{} \widetilde V_{\beta i}^{*} \widetilde V_{\alpha j}^{*} \widetilde V_{\beta j}^{}\right) 
\sin^2 \left(\frac{\widetilde\Delta_{ji}^{} L}{4 E}\right) \nonumber\\
&& + \sum_{i<j}^{} {\rm Im} \left( \widetilde V_{\alpha i}^{} \widetilde V_{\beta i}^{*} \widetilde V_{\alpha j}^{*} \widetilde V_{\beta j}^{}\right) 
\sin \left(\frac{\widetilde\Delta_{ji}^{} L}{2 E}\right)  
% (52)
\end{eqnarray}
with $i,j=1,2,3,4$.
Note that the results in Eq. (4) and Eqs. (39)--(52) only apply to a neutrino beam propagating in matter. When considering an antineutrino beam, we need to do the replacements $A \to -A$, $A^{\prime} \to -A^{\prime}$ and $V\to V^{*}$. Similar to the standard three-flavor mixing scheme, we discuss the $(3+1)$ flavor mixing in dense matter by considering the following four cases separately: 
case (NMO, $\nu_{3+1}^{}$),  case (IMO, $\nu_{3+1}^{}$),  case (NMO, $\overline\nu_{3+1}^{}$) and case (IMO, $\overline\nu_{3+1}^{}$). In the following discussion, $N_e^{}=N_n^{}=N_p^{}$ (i.e., $A^{\prime }=-A/2$ and $\epsilon \simeq 5 \times 10^{-5}$) is assumed and $V$ is 
parametrized as
\begin{eqnarray}
V=R_{34}^{} (\theta_{34}^{}, \delta_{34}) \cdot R_{24}^{} (\theta_{24}^{}, \delta_{24})\cdot
R_{14}^{} (\theta_{14}^{}, \delta_{14})\cdot U\;,
\end{eqnarray}
where $R_{ij}^{} (\theta_{ij}^{}, \delta_{ij}^{})$ (for $ij=13, 24,34$) represent the $4\times 4$ two-dimension rotation matrices
in the $(i, j)$ complex plane with the mixing angle $\theta_{ij}^{}$ and the CP-violating phase 
$\delta_{ij}^{}$, and $U$ has been defined in Eq. (16). Numerically, we typically take
the best-fit values of $(\theta_{12}^{}, \theta_{13}^{}, \theta_{23}^{}, \delta, \Delta_{21}^{}, \Delta_{31}^{})$ shown below Eq. (16), and $(\theta_{14}^{}, \theta_{24}^{},\theta_{34}^{})=(6.66^{\circ}, 7.81^{\circ}, 0^{\circ})$, $\delta_{14}^{}=\delta_{24}^{}=\delta_{34}^{}=0$ and $\Delta_{41}^{}=1.32~ {\rm eV^2}$ \cite{Diaz:2019fwt} for the active-sterile mixing part. Analytically, we expand the
exact expressions of $\widetilde \Delta_{ij}^{}$ (for $ij=21,31,41$) and $|\widetilde V_{\alpha i}^{}|^2$ (for $\alpha=e,\mu,\tau,s$ and $i=1,2,3,4$) in Eqs. (42), (43) and (48) in terms of $\Delta_{21}^{}/A$, $\Delta_{31}^{}/A$, $\Delta_{41}^{}/A$ and $\epsilon$. Thus the analytical approximations in 
this section are only valid for the $A\gg\Delta_{41}^{}$ range.
\subsection{(NMO, $\nu_{3+1}^{}$)}
Considering a neutrino beam with inverted mass ordering in the $(3+1)$ flavor mixing scheme,  we illustrate how $\widetilde \Delta_{ij}^{}$ (for $ij=21,31,41$) and $|\widetilde V_{\alpha i}^{}|^2$ (for $\alpha=e,\mu,\tau,s$ and $i=1,2,3,4$) evolve with the matter effect parameter $A$ in the upper left panel of Fig. 7 and Fig. 8, respectively. 
%(the evolutions of $\widetilde \Delta_{ij}^{}$ (for $ij=21,31,41$) with $A$ in cases \RNum{6}--\RNum{8} are also illustrated in Fig. 7, which we will discuss later.)
Note that we always have 
$\widetilde \Delta_{41}^{} > \widetilde \Delta_{31}^{} >\widetilde \Delta_{21}^{} $ in cases (NMO, $\nu_{3+1}^{}$) and (NMO, $\overline\nu_{3+1}^{}$), implying that there is no
intersection in  the upper row of Fig. 7.   
Comparing  Fig. 2 and Fig. 7, 
the evolutions of $|\widetilde V_{\alpha i}^{}|^2$ (for $\alpha=e,\mu,\tau$ and $i=1,2,3$) with $A$ in the $(3+1)$ flavor mixing scheme are similar to those of $|\widetilde U_{\alpha i}^{}|^2$ (for $\alpha=e,\mu,\tau$ and $i=1,2,3$) in the standard three-flavor mixing scheme if $A$ is not big enough (for example, $A<10^{-2}~{\rm eV}^2$). And it is the same case for $\widetilde \Delta_{ij}^{}$ (for $ij=21,31$).  However, if the matter density is very big, the neutrino flavor mixing in $(3+1)$ flavor mixing scheme can be very different from that in the standard three-flavor mixing scheme discussed in the last section.
By making perturbative expansions of Eq. (42) in terms of $\Delta_{21}^{}/A$, $\Delta_{31}^{}/A$, $\Delta_{41}^{}/A$ and $\epsilon$, we get
\vspace{1cm}
%%%%%%%%%%%%%%%%%% figure 7 %%%%%%%%%%%%%%%%%%%%%%%
\begin{figure}[h]
	\centering{
		\includegraphics[]{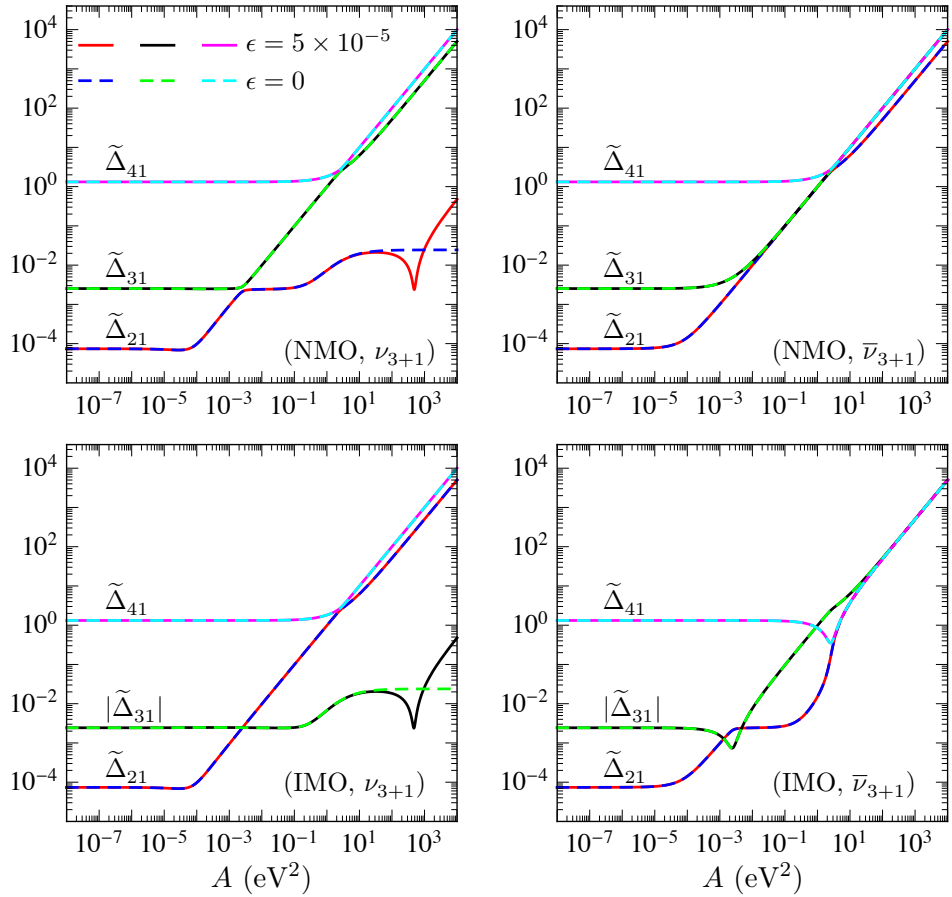}}
	\caption{In the $(3+1)$ mixing scheme, the illustration of  how the effective neutrino mass-squared differences $\widetilde \Delta_{21}^{}$, $|\widetilde \Delta_{31}^{}|$ and $\widetilde \Delta_{41}^{}$ evolve with the matter effect parameter $A$ in the
		case with or without radiative corrections, where we typically take the best-fit values of 
		$(\theta_{12}^{}, \theta_{13}^{}, \theta_{23}^{}, \delta,\Delta_{21}^{}, \Delta_{31}^{} )$ \cite{Esteban:2018azc}, and for the active-sterile neutrino mixing part  $(\theta_{14}^{}, \theta_{24}^{},\theta_{34}^{})=(6.66^{\circ}, 7.81^{\circ}, 0)$, $\delta_{14}^{}=\delta_{24}^{}=\delta_{34}^{}=0$, $\Delta_{41}^{}=1.32~ {\rm eV^2}$ \cite{Diaz:2019fwt}.}
\end{figure}
%%%%%%%%%%%%%%%%%%%%%%%%%%%%%%%%%%%%%%%%%%%%%%%%%%%
%%%%%%%%%%%%%%%%%% figure 8  %%%%%%%%%%%%%%%%%%%%%%%
\begin{figure}[h]
	\centering{
		\includegraphics[width=16cm]{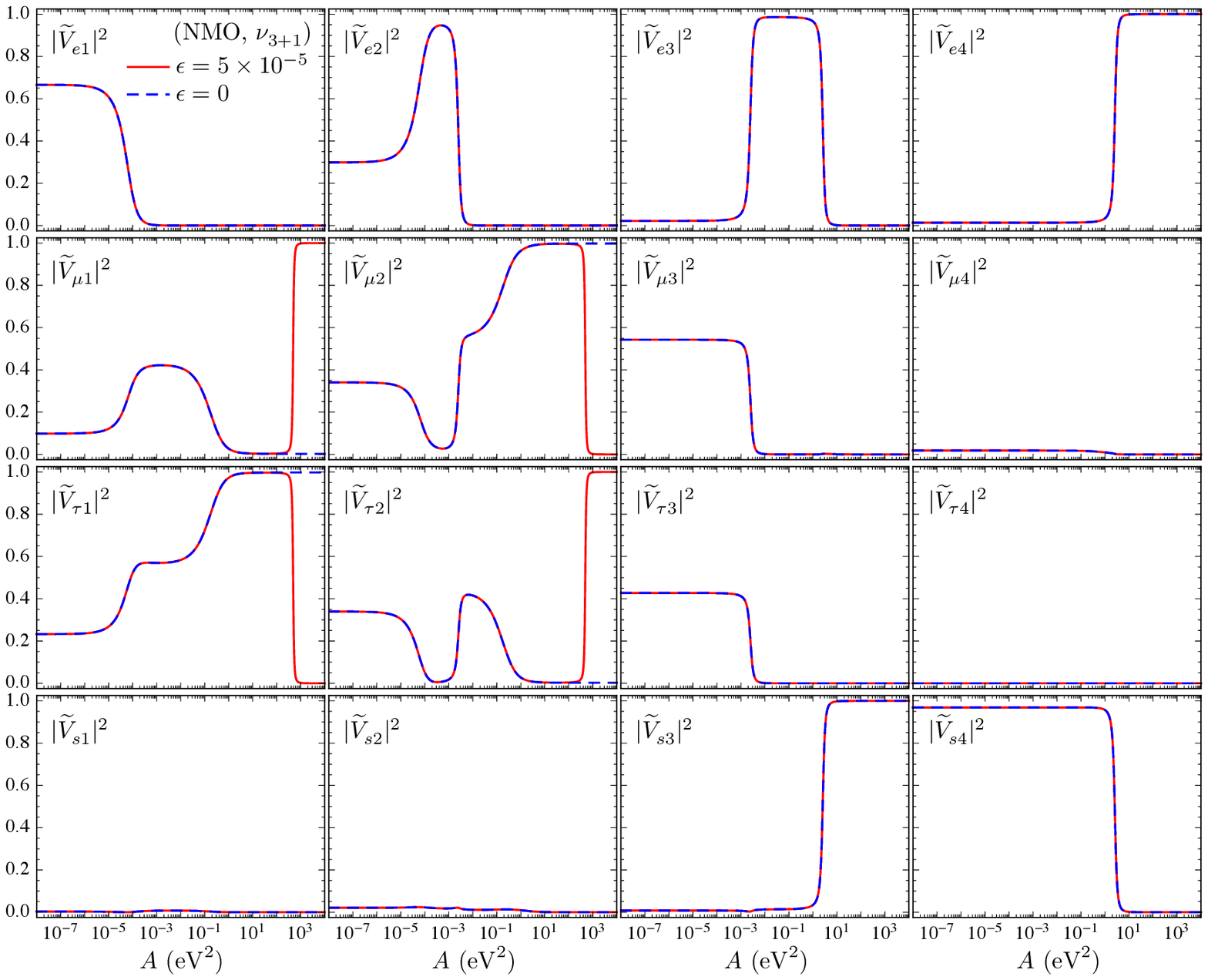}}
	\caption{In the $(3+1)$ mixing scheme, the illustration of  how $|\widetilde V_{\alpha i}^{}|^2$
		(for $\alpha=e,\mu,\tau,s~ {\rm and} ~i=1,2,3,4$) evolve with the matter effect parameter $A$ in the
		case $({\rm NMO}, \nu_{3+1}^{})$ with or without radiative corrections, where we typically take the best-fit values of 
		$(\theta_{12}^{}, \theta_{13}^{}, \theta_{23}^{}, \delta,\Delta_{21}^{}, \Delta_{31}^{} )$ \cite{Esteban:2018azc}, and for the active-sterile neutrino mixing part  $(\theta_{14}^{}, \theta_{24}^{},\theta_{34}^{})=(6.66^{\circ}, 7.81^{\circ}, 0)$, $\delta_{14}^{}=\delta_{24}^{}=\delta_{34}^{}=0$, $\Delta_{41}^{}=1.32~ {\rm eV^2}$ \cite{Diaz:2019fwt}.}
\end{figure}
%%%%%%%%%%%%%%%%%%%%%%%%%%%%%%%%%%%%%%%%%%%%%%%%%%%
\begin{eqnarray}
\widetilde\Delta_{21}^{} &\simeq& \xi_{s}^{} \nonumber\;,\\
\widetilde\Delta_{31}^{} &\simeq& \frac{A}{2}  -\frac{1}{2} \left[ A \epsilon - \xi_{s}^{} 
+ \left(1 - |V_{e2}^{}|^2 - 3 |V_{s2}^{}|^2\right) \Delta_{21}^{}
+ \left(1 - |V_{e3}^{}|^2 - 3 |V_{s3}^{}|^2\right) \Delta_{31}^{} \right.\nonumber\\&&\left.
+ \left(1 - |V_{e4}^{}|^2 - 3 |V_{s4}^{}|^2\right) \Delta_{41}^{}\right] \nonumber\;,\\
\widetilde\Delta_{41}^{} &\simeq& A - \frac{1}{2} \left[  A \epsilon - \xi_{s}^{} 
+ \left(1 -3  |V_{e2}^{}|^2 - |V_{s2}^{}|^2\right) \Delta_{21}^{}
+ \left(1 -3  |V_{e3}^{}|^2 -  |V_{s3}^{}|^2\right) \Delta_{31}^{} \right.\nonumber\\&&\left.
+\left(1 -3  |V_{e4}^{}|^2 -  |V_{s4}^{}|^2\right) \Delta_{41}^{}\right]\;,
%                 (54)
\end{eqnarray}
where
\begin{eqnarray}
\xi_{s}^{}&=&
\Big\{\left[ \left(|V_{\mu 2}^{}|^2 + |V_{\tau 2}^{}|^2 \right) \Delta_{21}^{}
+ \left(|V_{\mu 3}^{}|^2 + |V_{\tau 3}^{}|^2 \right) \Delta_{31}^{}
+ \left(|V_{\mu 4}^{}|^2 + |V_{\tau 4}^{}|^2 \right) \Delta_{41}^{} + A \epsilon\right]^2 
\nonumber\\&&
-4 A\epsilon \left( |V_{\mu 2}^{}|^2 \Delta_{21}^{} 
+|V_{\mu 3}^{}|^2 \Delta_{31}^{}  + |V_{\mu 4}^{}|^2\Delta_{41}^{} \right)
\nonumber\\&& 
- 4 \left(\Delta_{21}^{} \Delta_{31}^{}   C_{es}^{14}
+ \Delta_{21}^{} \Delta_{41}^{}  C_{es}^{13} +  \Delta_{31}^{} \Delta_{41}^{}   C_{es}^{12}\right)
\Big\}^{1/2}  \;.
%           (55)
\end{eqnarray}
From Eqs. (54) and (55), one can see that there are significant radiative corrections to $\widetilde \Delta_{21}^{}$
for the appearance of $A\epsilon$ in the leading order. Thus $\widetilde \Delta_{21}^{}$ will
approximately approach $A\epsilon$ instead of a constant value if $A$ is big enough.
We also notice that $\widetilde\Delta_{21}^{}$ has a minimum value by observing the
quadratic function of $A$ inside the brace of Eq. (55). The corresponding expression of $A$ is
\begin{eqnarray}
A &\simeq& \frac{1}{\epsilon} \left[\left( |V_{\mu 2}^{}|^2 -  |V_{\tau 2}^{}|^2\right) \Delta_{21}^{}
+ \left( |V_{\mu 3}^{}|^2 -  |V_{\tau 3}^{}|^2\right) \Delta_{31}^{} 
+\left( |V_{\mu 4}^{}|^2 -  |V_{\tau 4}^{}|^2\right) \Delta_{41}^{}
\right]\;
%                    (56)
\end{eqnarray}
which can be simplified as $A \simeq \left( |V_{\mu 4}^{}|^2 -  |V_{\tau 4}^{}|^2\right) \Delta_{41}^{}/\epsilon \simeq 481 ~{\rm eV}^2$ by
considering $\Delta_{41}^{}\gg \Delta_{31}^{}\gg\Delta_{21}^{}$ and the numerical input in Fig. 7.

By expanding Eq.  (48) in terms of $\Delta_{21}^{}/A$, $\Delta_{31}^{}/A$, $\Delta_{41}^{}/A$ and $\epsilon$, we approximately express $|\widetilde V_{\alpha i}^{}|^2$ as
\begin{eqnarray}
|\widetilde{V}_{e1}^{}|^2   \simeq  |\widetilde{V}_{e2}^{}|^2  &\simeq & |\widetilde{V}_{e3}^{}|^2 \simeq
0 \; , 
\quad |\widetilde{V}_{s1}^{}|^2   \simeq  |\widetilde{V}_{s2}^{}|^2  \simeq  |\widetilde{V}_{s4}^{}|^2 \simeq
0 \; , 
\quad |\widetilde{V}_{e4}^{}|^2 \simeq |\widetilde{V}_{s3}^{}|^2 \simeq 1 \; ,
\nonumber \\
|\widetilde{V}_{\mu 1}^{}|^2   \simeq |\widetilde{V}_{\tau 2}^{}|^2 &\simeq&
\frac{1}{2} +\frac{1}{2 \xi_{s}^{}}\left[A \epsilon+ \Delta_{21}^{} \left(
|V_{\tau 2}^{}|^2 - |V_{\mu 2}^{}|^2 \right) +
\Delta_{31}^{} \left(|V_{\tau 3}^{}|^2 - |V_{\mu 3}^{}|^2
\right) \right. \nonumber\\&&\left.+
\Delta_{41}^{} \left(|V_{\tau 4}^{}|^2 - |V_{\mu 4}^{}|^2
\right)\right] \; ,
\nonumber\\
|\widetilde{V}_{\mu 2}^{}|^2   \simeq  |\widetilde{V}_{\tau 1}^{}|^2 &\simeq&
\frac{1}{2} - \frac{1}{2 \xi_{s}^{}}\left[A \epsilon+ \Delta_{21}^{} \left(
|V_{\tau 2}^{}|^2 - |V_{\mu 2}^{}|^2 \right) +
\Delta_{31}^{} \left(|V_{\tau 3}^{}|^2 - |V_{\mu 3}^{}|^2
\right) \right.\nonumber\\&&\left.+
\Delta_{41}^{} \left(|V_{\tau 4}^{}|^2 - |V_{\mu 4}^{}|^2
\right)\right] \; ,\nonumber\\
|\widetilde{V}_{\mu 3}^{}|^2   \simeq  |\widetilde{V}_{\mu 4}^{}|^2 & \simeq&  |\widetilde{V}_{\tau 3}^{}|^2  \simeq  |\widetilde{V}_{\tau 4}^{}|^2\simeq 0 \;.
%                   （57）
\end{eqnarray}
Namely, the neutrino flavor mixing matrix $\widetilde V$
can be approximately described by one degree of freedom,
\begin{eqnarray}
\left. \widetilde{V}\right|_{A\gg\Delta_{41}^{}} \simeq
\begin{pmatrix} 0 & 0 & 0& 1 \cr 
\cos\theta_s^{} & ~\sin\theta_s^{}~ & 0 &0\cr
-\sin\theta_s^{} & \cos\theta_s^{} & 0& 0 \cr
0&0&1&0
\end{pmatrix} \; ,
%     (58)
\end{eqnarray}
where $\theta_s^{} \in [0, \pi/2]$ and
$\tan^2\theta_s^{}\simeq|\widetilde V_{\mu 2}^{}|^2 /|\widetilde V_{\mu 1}^{}|^2 $ with 
$|\widetilde V_{\mu 1}^{}|^2$ and $|\widetilde V_{\mu 2}^{}|^2$ being taken from Eq. (57).
The corresponding neutrino oscillation probabilities $\widetilde P_{\alpha\beta}^{}$ (for $\alpha \beta =e,\mu,\tau,s$) in Eq. (52) are reduced to
\begin{align}
{\widetilde P}_{ee}^{}&\simeq 1,  &{\widetilde P}_{e\mu}^{}&\simeq 0, &{\widetilde P}_{e\tau}^{}&\simeq 0, 
&{\widetilde P}_{es}^{}&\simeq 0 \;,
\nonumber\\
{\widetilde P}_{\mu e}^{} &\simeq 0,  &{\widetilde P}_{\mu\mu}^{} &\simeq 1 - \sin^2 2 \theta_s^{} \sin^2 \frac{\widetilde\Delta_{21}L}{4 E}, &{\widetilde P}_{\mu\tau}^{}&\simeq \sin^2 2 \theta_s^{} \sin^2 \frac{\widetilde\Delta_{21}L}{4 E}, 
&{\widetilde P}_{\mu s}^{}&\simeq 0 \;,
\nonumber\\
{\widetilde P}_{\tau e}^{} &\simeq 0,  &{\widetilde P}_{\tau\mu}^{} &\simeq  \sin^2 2 \theta_s^{} \sin^2 \frac{\widetilde\Delta_{21}L}{4 E}, &{\widetilde P}_{\tau\tau}^{} &\simeq 1 - \sin^2 2 \theta_s^{} \sin^2 \frac{\widetilde\Delta_{21}L}{4 E}, 
&{\widetilde P}_{\tau s}^{}&\simeq 0  \;,\nonumber\\
{\widetilde P}_{se}^{}& \simeq 0,  &{\widetilde P}_{s\mu}^{}&\simeq 0, &{\widetilde P}_{s\tau}^{}&\simeq 0, 
&{\widetilde P}_{ss}^{}&\simeq 1\;,
\nonumber\\
%                 (59)
\end{align}
where $\widetilde \Delta_{21}^{}$ has been given in Eq. (54) and $\sin^2 2\theta_s^{} = |\widetilde V_{\mu 1}^{} |^2
(1- |\widetilde V_{\mu 1}^{} |^2)$ with $ |\widetilde V_{\mu 1}^{} |^2$ being derived in Eq. (57).  The absolute errors
of $\widetilde P_{\alpha\beta}^{}$ are illustrated in Fig. 9, from which
we can see that Eq. (59) is a good approximation in a wide range of $L/E$ and $A/\Delta_{41}^{}$. For the
upper left part in each subgraph of Fig. 9, the accuracies are not good enough and the largest errors
in $\Delta \widetilde P_{\mu\mu}^{}$, $\Delta \widetilde P_{\mu\tau}^{}$, $\Delta \widetilde P_{\tau\mu}^{}$ and $\Delta \widetilde P_{\tau\tau}^{}$ appear around the minimum of $\widetilde \Delta_{21}^{}$ ($A/\Delta_{41}\sim 4 \times 10^2$). This can be improved by
keeping higher orders of $\Delta_{21}/A$, $\Delta_{31}/A$, $\Delta_{41}/A$ and $\epsilon$ or
just making perturbative expansions in terms of $\Delta_{21}/A$, $\Delta_{31}/A$ and $\epsilon$.

%%%%%%%%%%%%%%%%%% figure 9 %%%%%%%%%%%%%%%%%%%%%%%
\begin{figure}[H]
	\centering{
		\includegraphics[width=16cm]{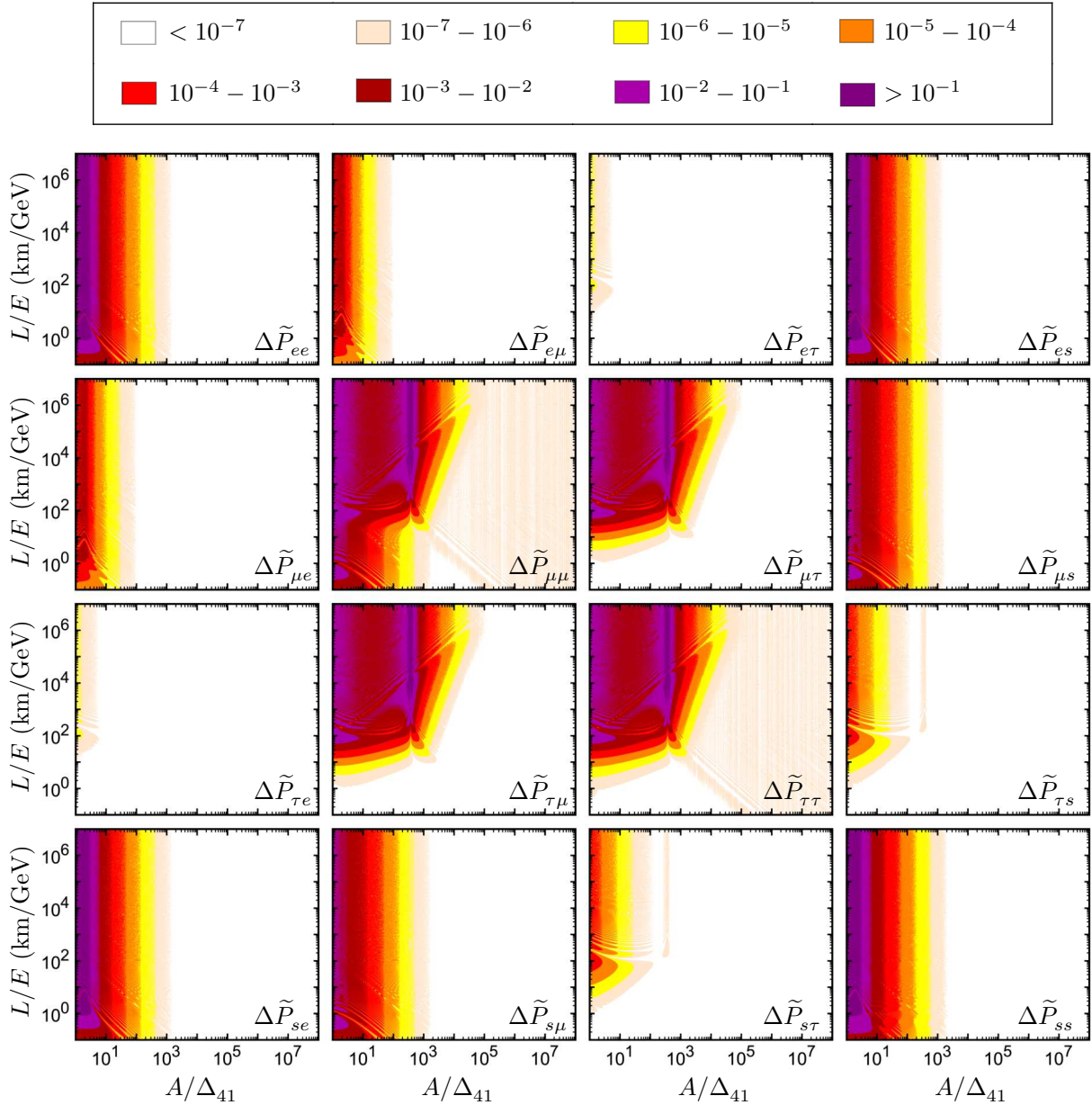}}
	\caption{The absolute errors of our analytical approximations in Eq. (59), where 
		the best-fit values of 
		$(\theta_{12}^{}, \theta_{13}^{}, \theta_{23}^{}, \delta,\Delta_{21}^{}, \Delta_{31}^{} )$ in Ref. \cite{Esteban:2018azc}  have been input,
		$(\theta_{14}^{}, \theta_{24}^{},\theta_{34}^{})=(6.66^{\circ}, 7.81^{\circ}, 0)$, $\delta_{14}^{}=\delta_{24}^{}=\delta_{34}^{}=0$, $\Delta_{41}^{}=1.32~ {\rm eV^2}$ \cite{Diaz:2019fwt} and  $\epsilon\simeq 5 \times 10^{-5}$ \cite{Botella:1986wy}.}
\end{figure}

Specifically, if $A\epsilon$ is negligible in Eqs. (54) and (57), one may abandon the smaller terms
of $\Delta_{21}^{}$ and $\Delta_{31}^{}$, and get
\begin{eqnarray}
\widetilde \Delta_{21}^{} &\simeq&  \left(|V_{\mu 4}^{}|^2  + |V_{\tau 4}^{}|^2\right) \Delta_{41}^{}\;,
\nonumber\\  
|\widetilde V_{\mu 1}^{}|^2 &\simeq& |\widetilde V_{\tau 2}^{}|^2 \simeq \frac{|V_{\tau 4}^{}|^2}{ |V_{\mu 4}^{}|^2 + |V_{\tau 4}^{}|^2 }
\;, 
\nonumber\\    |\widetilde V_{\mu 2}^{}|^2 &\simeq&  |\widetilde V_{\tau 1}^{}|^2 \simeq\frac{|V_{\mu 4}^{}|^2}{ |V_{\mu 4}^{}|^2 + |V_{\tau 4}^{}|^2 }\;,
%             (60)
\end{eqnarray}
where $|V_{\mu 4}^{}|^2=\cos^2 \theta_{14}^{} \sin^2 \theta_{24}$, $|V_{\tau 4}^{}|^2=\cos^2 \theta_{14}^{} \cos^2 \theta_{24}^{}\sin^2 \theta_{34}$ from the parametrization of $V$ in Eq. (53).
This is equivalent to the asymptotic behaviors of $\widetilde \Delta_{21}^{}$ and $
|\widetilde V_{\alpha i}^{}|^2$ (for $\alpha i =\mu 1,\mu 2,\tau 1,\tau 2$) in very dense matter in the case without radiative corrections (i.e. the blue dashed lines in Fig. 8). Due to the typical value $\theta_{34}^{}=0$ inputted in Fig. 8,
we get $|\widetilde V_{\mu 1}^{}|^2 \simeq |\widetilde V_{\tau 2}^{}|^2 \simeq 0$ and $
|\widetilde V_{\mu 2}^{}|^2 \simeq |\widetilde V_{\tau 1}^{}|^2 \simeq 1$. By choosing a non-zero value of $\theta_{34}^{}$, the neutrino flavor mixing can be very different. With the increase of $A$,
the $A \epsilon$ term in Eqs. (54) and (57) will become dominate. In the $  A \to \infty$ limit,
$|\widetilde V_{\mu 1}^{}|^2 \simeq |\widetilde V_{\tau 2}^{}|^2 \simeq 1$ and $
|\widetilde V_{\mu 2}^{}|^2 \simeq |\widetilde V_{\tau 1}^{}|^2 \simeq 0$ can be derived from Eq. (57).
Considering this extreme case, we summarize the corresponding analytical expressions of $\widetilde \Delta_{ji}^{}$ (for $ji=21,31, 41$) and $\widetilde V$ in Table 2, where the four cases (NMO, $\nu_{3+1}^{}$),  (IMO, $\nu_{3+1}^{}$),  (NMO, $\overline\nu_{3+1}^{}$) and (IMO, $\overline\nu_{3+1}^{}$) are all considered separately.
With $|\widetilde V_{\alpha i}^{}|^2$ taking 0 or 1 (or $\theta_s^{}\to 0$ or $\pi/2$), there will be no neutrino oscillation between the four flavors and it makes no sense to discuss lepton flavor mixing.
%%%%%%%%%%%%%%%%%%%%%   table  2  %%%%%%%%%%%%%%%%%%%%%%%%%%%
%%%%%%%%%%%%%%%%%%%%%%%%%
\begin{table}[h]
	\caption{In the $(3+1)$ mixing scheme, the analytical expressions of $\widetilde \Delta_{ji}^{}$ (for $ji=21,31, 41$) 
		and $\widetilde V$ in the $A\to \infty$ limit, where we only show the terms of  order ${\cal O} (A)$ for $\widetilde \Delta_{ji}^{}$.}
	\vspace{0.1cm}
	\centering
	\renewcommand\arraystretch{1.3}
	\begin{tabular}{c|c|c|c|c} \hline\hline
		&(NMO, $\nu_{3+1}^{}$)&(IMO, $\nu_{3+1}^{}$)&(NMO, $\overline\nu_{3+1}^{}$)&(IMO, $\overline\nu_{3+1}^{}$)\\\hline
		$\widetilde \Delta_{21}^{}$& $A\epsilon$&$ A/2-A\epsilon$&$A/2$&$A/2-A\epsilon$\\	\hline
		$\widetilde \Delta_{31}^{}$&$A/2$& $-A\epsilon$&$A-A\epsilon$&$-A/2$\\\hline
		$\widetilde \Delta_{41}^{}$&$A$& $A-A\epsilon$&$A$&$A/2$\\\hline
		$\widetilde V $&
		$\begin{pmatrix}
		0&0&0&1\cr
		1&0&0&0\cr
		0&1&0&0\cr
		0&0&1&0\cr
		\end{pmatrix}$&$\begin{pmatrix}
		0&0&0&1\cr
		0&0&1&0\cr
		1&0&0&0\cr
		0&1&0&0\cr
		\end{pmatrix}$&	$\begin{pmatrix}
		1&0&0&0\cr
		0&0&0&1\cr
		0&0&1&0\cr
		0&1&0&0\cr
		\end{pmatrix}$&	$\begin{pmatrix}
		0&0&1&0\cr
		0&0&0&1\cr
		0&1&0&0\cr
		1&0&0&0\cr
		\end{pmatrix}$\\\hline\hline
	\end{tabular}
\end{table}
%%%%%%%%%%%%%%%%%%%%%%%%%%%%%%%%%%%%%%%%%%%%%%%%%%%%%%%%%%%%%%%%%%%%%%%%%%%%%%%%%%%%%%%

\subsection{(IMO, $\nu^{}_{3+1}$)}
Given a neutrino beam with inverted mass ordering in the $(3+1)$ flavor mixing scheme,  the evolutions of $\widetilde \Delta_{ij}^{}$ (for $ij=21,31,41$)
and $|\widetilde V_{\alpha i}^{}|^2$ (for $\alpha=e, \mu,\tau,s$ and $i=1,2,3,4$) with $A$ are demonstrated
in the lower left panel of Fig. 7 and Fig. 10, respectively.
In this case, $\widetilde \Delta_{31}^{} < 0 <\widetilde \Delta_{21}^{} <\widetilde \Delta_{41}^{} $ always holds from Fig. 7. By making perturbative expansions of Eq. (43), we get 
\begin{eqnarray}
\widetilde\Delta_{21}^{} &\simeq& \frac{A}{2}  -\frac{1}{2} \left[ A \epsilon + \xi_{s}^{} 
+ \left(1 - |V_{e2}^{}|^2 - 3 |V_{s2}^{}|^2\right) \Delta_{21}^{}
+ \left(1 - |V_{e3}^{}|^2 - 3 |V_{s3}^{}|^2\right) \Delta_{31}^{} \right.\nonumber\\&&\left.
+ \left(1 - |V_{e4}^{}|^2 - 3 |V_{s4}^{}|^2\right) \Delta_{41}^{}\right] \nonumber\;,\\
\widetilde\Delta_{31}^{} &\simeq& - \xi_{s}^{} \nonumber\;,\\
\widetilde \Delta_{41}^{} &\simeq& A - \frac{1}{2} \left[  A \epsilon +\xi_{s}^{} 
+ \left(1 -3  |V_{e2}^{}|^2 - |V_{s2}^{}|^2\right) \Delta_{21}^{}
+ \left(1 -3  |V_{e3}^{}|^2 -  |V_{s3}^{}|^2\right) \Delta_{31}^{} \right.\nonumber\\&&\left.
+\left(1 -3  |V_{e4}^{}|^2 -  |V_{s4}^{}|^2\right) \Delta_{41}^{}\right] \;,
%                 (61)
\end{eqnarray}
where $ \xi_{s}^{} $ has been defined in Eq. (55). From  Eq. (61), it is easy to see
that there are significant radiative corrections to $\widetilde \Delta_{31}^{}$ with the $\epsilon$
term appearing in the leading order. 
There are a minimum of $\widetilde\Delta_{31}^{}$ corresponding to $A$ in Eq. (56).
Similarly, Eq. (48) can be reduced to
%%%%%%%%%%%%%%%%%%%%%%%%%%%%%%%%%%%%%%%%%%%%%%%%%%%
%%%%%%%%%%%%%%%%%%  figure 10 %%%%%%%%%%%%%%%%%%%%%%%
\begin{figure}[h]
	\centering{
		\includegraphics[width=16cm]{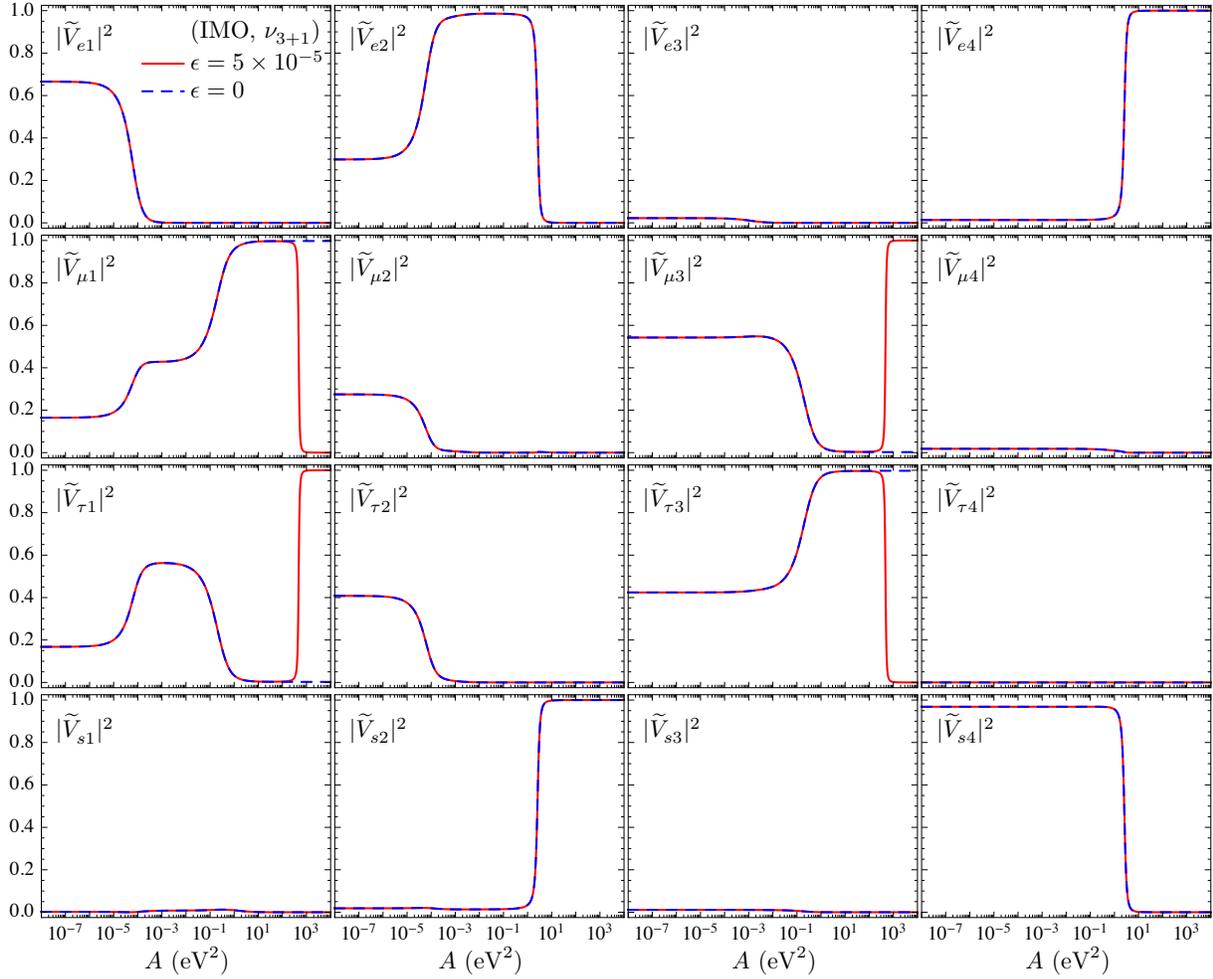}}
	\caption{In the $(3+1)$ mixing scheme, the illustration of  how $|\widetilde V_{\alpha i}^{}|^2$
		(for $\alpha=e,\mu,\tau,s~ {\rm and} ~i=1,2,3,4$) evolve with the matter effect parameter $A$ in the
		case $({\rm IMO}, \nu_{3+1}^{})$ with or without radiative corrections, where we typically take the best-fit values of 
		$(\theta_{12}^{}, \theta_{13}^{}, \theta_{23}^{}, \delta,\Delta_{21}^{}, \Delta_{31}^{} )$ \cite{Esteban:2018azc}, and for the active-sterile neutrino mixing part  $(\theta_{14}^{}, \theta_{24}^{},\theta_{34}^{})=(6.66^{\circ}, 7.81^{\circ}, 0)$, $\delta_{14}^{}=\delta_{24}^{}=\delta_{34}^{}=0$, $\Delta_{41}^{}=1.32~ {\rm eV^2}$ \cite{Diaz:2019fwt}.}
\end{figure}
%%%%%%%%%%%%%%%%%%%%%%%%%%%%%%%%%%%%%%%%%%%%%%%%%%%
\begin{eqnarray}
|\widetilde{V}_{e1}^{}|^2   \simeq  |\widetilde{V}_{e2}^{}|^2  &\simeq & |\widetilde{V}_{e3}^{}|^2 \simeq
0 \; , 
\quad |\widetilde{V}_{s1}^{}|^2   \simeq  |\widetilde{V}_{s3}^{}|^2  \simeq  |\widetilde{V}_{s4}^{}|^2 \simeq
0 \; , 
\quad |\widetilde{V}_{e4}^{}|^2 \simeq |\widetilde{V}_{s2}^{}|^2 \simeq 1 \; ,
\nonumber \\
|\widetilde{V}_{\mu 1}^{}|^2   \simeq |\widetilde{V}_{\tau 3}^{}|^2 &\simeq&
\frac{1}{2} - \frac{1}{2 \xi_{s}^{}}\left[A \epsilon+ \Delta_{21}^{} \left(
|V_{\tau 2}^{}|^2 - |V_{\mu 2}^{}|^2 \right) +
\Delta_{31}^{} \left(|V_{\tau 3}^{}|^2 - |V_{\mu 3}^{}|^2
\right) \right.\nonumber\\&&\left.+
\Delta_{41}^{} \left(|V_{\tau 4}^{}|^2 - |V_{\mu 4}^{}|^2
\right)\right] \; ,
\nonumber\\
|\widetilde{V}_{\mu 3}^{}|^2   \simeq  |\widetilde{V}_{\tau 1}^{}|^2 &\simeq&
\frac{1}{2} +\frac{1}{2 \xi_{s}^{}}\left[A \epsilon+ \Delta_{21}^{} \left(
|V_{\tau 2}^{}|^2 - |V_{\mu 2}^{}|^2 \right) +
\Delta_{31}^{} \left(|V_{\tau 3}^{}|^2 - |V_{\mu 3}^{}|^2
\right) \right.\nonumber\\&&\left.+
\Delta_{41}^{} \left(|V_{\tau 4}^{}|^2 - |V_{\mu 4}^{}|^2
\right)\right] \; ,\nonumber\\
|\widetilde{V}_{\mu 2}^{}|^2   \simeq  |\widetilde{V}_{\mu 4}^{}|^2 & \simeq&  |\widetilde{V}_{\tau 2}^{}|^2  \simeq  |\widetilde{V}_{\tau 4}^{}|^2\simeq 0 \;.
%                  （62）
\end{eqnarray}
This means $\widetilde V$ can be approximately parametrized as
\begin{eqnarray}
\left. \widetilde{V}\right|_{A\gg\Delta_{41}^{}} \simeq
\begin{pmatrix} 0 & 0 & 0& 1 \cr 
\cos\theta_s^{} &0& \sin\theta_s^{}&0 \cr
-\sin\theta_s^{} &0& \cos\theta_s^{} & 0 \cr
0&1&0&0
\end{pmatrix} \; ,
%     (63)
\end{eqnarray}
where $\theta_s^{} \in [0, \pi/2]$ and $\tan^2\theta_s^{}=|\widetilde V_{\mu 3}^{}|^2/ |\widetilde V_{\mu 1}^{}|^2 $ with $|\widetilde V_{\mu 1}^{}|^2$ and  $|\widetilde V_{\mu 3}^{}|^2$ having been shown in Eq. (62).
One can derive the corresponding approximate neutrino oscillation probability by substituting $\widetilde \Delta_{31}^{}$ for $\widetilde \Delta_{21}^{}$ in Eq. (59).
Similar to the standard three-flavor mixing scheme, it is impossible to discriminate between the normal
mass ordering and the inverted mass ordering from neutrino oscillations in the $(3+1)$ flavor mixing scheme if the matter density is very big.

Note that if the $A\epsilon$ term  in Eqs. (61) and (62) is negligible, we omit the smaller terms of $\Delta_{21}^{}$ and $\Delta_{31}^{}$
and obtain
\begin{eqnarray}
\widetilde \Delta_{31}^{} &\simeq&  -\left(|V_{\mu 4}^{}|^2  + |V_{\tau 4}^{}|^2\right) \Delta_{41}^{}\;,
\nonumber\\  
|\widetilde V_{\mu 1}^{}|^2 &\simeq& |\widetilde V_{\tau 3}^{}|^2 \simeq \frac{|V_{\mu 4}^{}|^2}{ |V_{\mu 4}^{}|^2 + |V_{\tau 4}^{}|^2 }
\;, 
\nonumber\\    |\widetilde V_{\mu 3}^{}|^2 &\simeq&  |\widetilde V_{\tau 1}^{}|^2 \simeq\frac{|V_{\tau 4}^{}|^2}{ |V_{\mu 4}^{}|^2 + |V_{\tau 4}^{}|^2 }\;,
%             (64)
\end{eqnarray}
which is consistent with the asymptotic behaviors of $\widetilde \Delta_{31}^{}$ and $
|\widetilde V_{\alpha i}^{}|^2$ (for $\alpha i =\mu 1,\mu 3,\tau 1,\tau 3$)  in very dense matter in the case without radiative corrections (i.e. the blue dashed line in Fig. 10).
By inputting $\theta_{34}^{} =0$ (i.e., $|V_{\tau 4}^{}|^2 =0$), we directly get $|\widetilde V_{\mu 1}^{}|^2 \simeq |\widetilde V_{\tau 3}^{}|^2 \simeq1$ and $|\widetilde V_{\mu 3}^{}|^2 \simeq |\widetilde V_{\tau 1}^{}|^2 \simeq 0$ from Eq. (64).
If the limit of $A \to \infty$ is taken, 
 one can infer from Eq. (62) that $|\widetilde V_{\mu 1}^{}|^2 \simeq |\widetilde V_{\tau 3}^{}|^2 \simeq 0$ and $|\widetilde V_{\mu 3}^{}|^2 \simeq |\widetilde V_{\tau 1}^{}|^2 \simeq 1$, leading to no neutrino oscillations. 
\subsection{(NMO, $\overline\nu_{3+1}^{}$)}
When it comes to a neutrino beam with normal mass ordering in the (3+1) flavor mixing scheme,
the corresponding evolutions of $\widetilde \Delta_{ij}^{}$ (for $ij=21,31,41$) and $|\widetilde V_{\alpha i}^{}|^2$ (for $\alpha =e,\mu,\tau,s$ and $i=1,2,3,4$) with the matter effect parameter $A$ are illustrated in the upper right panel of Fig. 7 and Fig. 11, respectively. Analytically, we perform perturbative expansions of $\widetilde \Delta_{ij}^{}$ (for $ij=21,31,41$) in
Eq. (42)  and get
%%%%%%%%%%%%%%%%%% figure 11  %%%%%%%%%%%%%%%%%%%%%%%
\begin{figure}[h]
	\centering{
		\includegraphics[width=16cm]{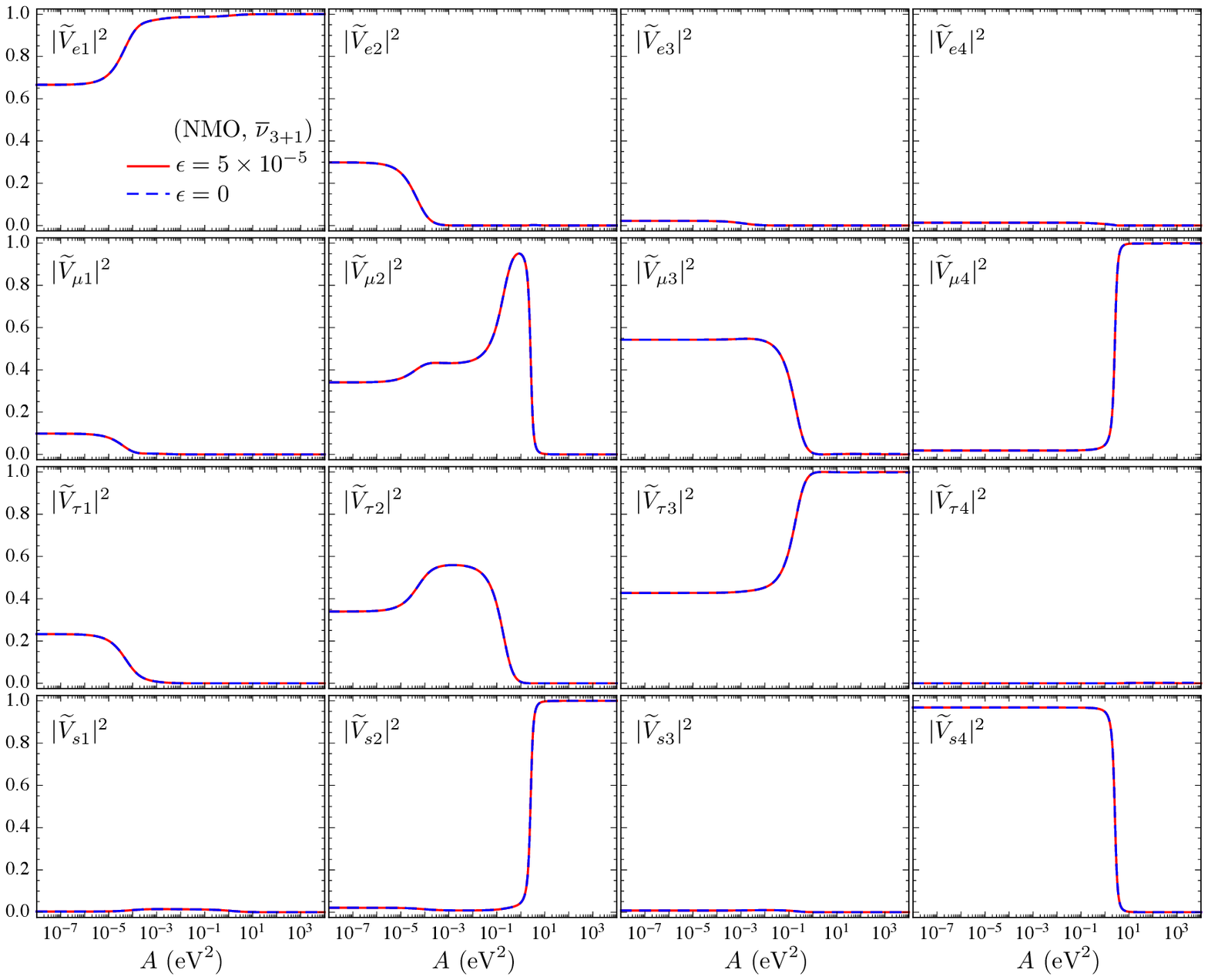}}
	\caption{In the $(3+1)$ mixing scheme, the illustration of  how $|\widetilde V_{\alpha i}^{}|^2$
		(for $\alpha=e,\mu,\tau,s~ {\rm and} ~i=1,2,3,4$) evolve with the matter effect parameter $A$ in the
		case $({\rm NMO},\overline \nu_{3+1}^{})$ with or without radiative corrections, where we typically take the best-fit values of 
		$(\theta_{12}^{}, \theta_{13}^{}, \theta_{23}^{}, \delta,\Delta_{21}^{}, \Delta_{31}^{} )$ \cite{Esteban:2018azc}, and for the active-sterile neutrino mixing part  $(\theta_{14}^{}, \theta_{24}^{},\theta_{34}^{})=(6.66^{\circ}, 7.81^{\circ}, 0)$, $\delta_{14}^{}=\delta_{24}^{}=\delta_{34}^{}=0$, $\Delta_{41}^{}=1.32~ {\rm eV^2}$ \cite{Diaz:2019fwt}.}
\end{figure}
%%%%%%%%%%%%%%%%%%%%%%%%%%%%%%%%%%%%%%%%%%%%%%%%%%%
\begin{eqnarray}
\widetilde\Delta_{21}^{} &\simeq& \frac{A}{2}  -\frac{1}{2} \left[ 
\left(|V_{e2}^{}|^2 - |V_{s2}^{}|^2\right) \Delta_{21}^{}
+ \left( |V_{e3}^{}|^2 - |V_{s3}^{}|^2\right) \Delta_{31}^{}
+ \left( |V_{e4}^{}|^2 -  |V_{s4}^{}|^2\right) \Delta_{41}^{}\right] \nonumber\;,\\ 
\widetilde\Delta_{31}^{} &\simeq& A  -\frac{1}{2} \left[ A \epsilon + \xi_{s}^{} 
- \left(1 -3  |V_{e2}^{}|^2 - |V_{s2}^{}|^2\right) \Delta_{21}^{}
- \left(1 -3  |V_{e3}^{}|^2 -  |V_{s3}^{}|^2\right) \Delta_{31}^{} \right.\nonumber\\&&\left.
-\left(1 -3  |V_{e4}^{}|^2 -  |V_{s4}^{}|^2\right) \Delta_{41}^{}\right] \nonumber\;,\\
\widetilde\Delta_{41}^{} &\simeq& A - \frac{1}{2} \left[  A \epsilon - \xi_{s}^{} 
- \left(1 -3  |V_{e2}^{}|^2 - |V_{s2}^{}|^2\right) \Delta_{21}^{}
-\left(1 -3  |V_{e3}^{}|^2 -  |V_{s3}^{}|^2\right) \Delta_{31}^{} \right.\nonumber\\&&\left.
-\left(1 -3  |V_{e4}^{}|^2 -  |V_{s4}^{}|^2\right) \Delta_{41}^{}\right] \;,
%             (65)
\end{eqnarray}
where
\begin{eqnarray}
\xi_{s}^{}&=&
\Big\{\left[ \left(|V_{\mu 2}^{}|^2 + |V_{\tau 2}^{}|^2 \right) \Delta_{21}^{}
+ \left(|V_{\mu 3}^{}|^2 + |V_{\tau 3}^{}|^2 \right) \Delta_{31}^{}
+ \left(|V_{\mu 4}^{}|^2 + |V_{\tau 4}^{}|^2 \right) \Delta_{41}^{} - A \epsilon\right]^2 
\nonumber\\&& 
+ 4 A\epsilon \left( |V_{\mu 2}^{}|^2\Delta_{21}^{} 
+|V_{\mu 3}^{}|^2 \Delta_{31}^{}  + |V_{\mu 4}^{}|^2\Delta_{41}^{} \right) \nonumber\\
&&- 4 \left(\Delta_{21}^{} \Delta_{31}^{}   C_{es}^{14}
+ \Delta_{21}^{} \Delta_{41}^{}  C_{es}^{13} +  \Delta_{31}^{} \Delta_{41}^{}   C_{es}^{12}\right)
\Big\}^{1/2} \;.
%               (66)
\end{eqnarray}
Note that there are no significant radiative corrections to $\widetilde\Delta_{21}^{}$, $\widetilde\Delta_{31}^{}$
and $\widetilde\Delta_{41}^{}$ with the $A\epsilon$ term appearing in the next-to-leading order or higher order.
Similarly, $|\widetilde V_{\alpha i}^{}|^2$ can be reduced to
\begin{eqnarray}
|\widetilde{V}_{e2}^{}|^2   \simeq  |\widetilde{V}_{e3}^{}|^2  &\simeq & |\widetilde{V}_{e4}^{}|^2 \simeq
0 \; , 
\quad |\widetilde{V}_{s1}^{}|^2   \simeq  |\widetilde{V}_{s3}^{}|^2  \simeq  |\widetilde{V}_{s4}^{}|^2 \simeq
0 \; , 
\quad |\widetilde{V}_{e1}^{}|^2 \simeq |\widetilde{V}_{s2}^{}|^2 \simeq 1 \; ,
\nonumber \\
|\widetilde{V}_{\mu 3}^{}|^2   \simeq |\widetilde{V}_{\tau 4}^{}|^2 &\simeq&
\frac{1}{2} +\frac{1}{2 \xi_{s}^{}}\left[-A \epsilon+ \Delta_{21}^{} \left(
|V_{\tau 2}^{}|^2 - |V_{\mu 2}^{}|^2 \right) +
\Delta_{31}^{} \left(|V_{\tau 3}^{}|^2 - |V_{\mu 3}^{}|^2
\right) \right.\nonumber\\&&\left.+
\Delta_{41}^{} \left(|V_{\tau 4}^{}|^2 - |V_{\mu 4}^{}|^2
\right)\right] \; ,
\nonumber\\
|\widetilde{V}_{\mu 4}^{}|^2   \simeq  |\widetilde{V}_{\tau 3}^{}|^2 &\simeq&
\frac{1}{2} - \frac{1}{2 \xi_{s}^{}}\left[-A \epsilon+ \Delta_{21}^{} \left(
|V_{\tau 2}^{}|^2 - |V_{\mu 2}^{}|^2 \right) +
\Delta_{31}^{} \left(|V_{\tau 3}^{}|^2 - |V_{\mu 3}^{}|^2
\right) \right.\nonumber\\&&\left.+
\Delta_{41}^{} \left(|V_{\tau 4}^{}|^2 - |V_{\mu 4}^{}|^2
\right)\right] \; ,\nonumber\\
|\widetilde{V}_{\mu 1}^{}|^2   \simeq  |\widetilde{V}_{\mu 2}^{}|^2 & \simeq&  |\widetilde{V}_{\tau 1}^{}|^2  \simeq  |\widetilde{V}_{\tau 2}^{}|^2\simeq 0 \;.
%             (67)
\end{eqnarray}
Namely,
\begin{eqnarray}
\left. \widetilde{V}\right|_{A\gg\Delta_{41}^{}} \simeq
\begin{pmatrix} 1 & 0 & 0& 0\cr 
0 &0&\cos\theta_s^{}& \sin\theta_s^{} \cr
0&0&-\sin\theta_s^{} & \cos\theta_s^{}  \cr
0&1&0&0
\end{pmatrix} \; ,
%     (68)
\end{eqnarray}
where $\theta_s^{} \in [0, \pi/2]$ and $\tan^2\theta_s^{}=|\widetilde V_{\mu 4}^{}|^2/ |\widetilde V_{\mu 3}^{}|^2 $ with $|\widetilde V_{\mu 3}^{}|^2$ and  $|\widetilde V_{\mu 4}^{}|^2$ being expressed in Eq. (67).
The corresponding approximate formulas of neutrino oscillation probabilities can be obtained by doing the replacement $\widetilde \Delta_{21}^{} \to \widetilde \Delta_{43}^{}$ in Eq. (59). And we have
$\widetilde \Delta_{43}^{} \simeq \xi_s^{}$ from Eq. (65).
If the $A\epsilon$ term is not dominant in Eq. (65), one can ignore the smaller terms of $A\epsilon$, $\Delta_{21}^{}$ and $\Delta_{31}^{}$ 
and get
\begin{eqnarray}
|\widetilde V_{\mu 3}^{}|^2 &\simeq& |\widetilde V_{\tau 4}^{}|^2 \simeq \frac{|V_{\tau 4}^{}|^2}{ |V_{\mu 4}^{}|^2 + |V_{\tau 4}^{}|^2 }
\;, \nonumber\\    
|\widetilde V_{\mu 4}^{}|^2 &\simeq&  |\widetilde V_{\tau 3}^{}|^2 \simeq\frac{|V_{\mu 4}^{}|^2}{ |V_{\mu 4}^{}|^2 + |V_{\tau 4}^{}|^2 } \;,
%             (69)
\end{eqnarray} 
which are equivalent to the fixed values of $|\widetilde V_{\alpha i}^{}|^2$ (for $\alpha i = \mu 3, \mu 4,\tau 3,\tau 4$) in the $A\to\infty$ limit when the radiative corrections are not included.
 By inputting $\theta_{34}^{}=0$  (i.e., $|V_{\tau 4}^{}|^2 =0$), Eq. (69) turns into $|\widetilde V_{\mu 3}^{}|^2 \simeq |\widetilde V_{\tau 4}^{}|^2 \simeq 0$ and $|\widetilde V_{\mu4}^{}|^2 \simeq |\widetilde V_{\tau 3}^{}|^2 \simeq 1$.
 In the $A\to \infty$ limit, we also have the same results: $|\widetilde V_{\mu 3}^{}|^2 \simeq |\widetilde V_{\tau 4}^{}|^2 \simeq 0$ and $|\widetilde V_{\mu4}^{}|^2 \simeq |\widetilde V_{\tau 3}^{}|^2 \simeq 1$ with $\xi_s^{} \to
 A \epsilon$  in Eq. (67). Thus there will be no neutrino
 oscillations in the $A \to \infty $ limit with $\theta_s^{}\simeq \arctan(|\widetilde V_{\mu 4}^{}|/|\widetilde V_{\mu 3}^{}|)  \simeq\pi/2 $.
Note that there is no distinguishable difference between the cases with or without 
 radiative corrections in Fig. 11. However, by inputting a non-zero value of $\theta_{34}^{}$, one may see significant
 radiative corrections to neutrino flavor mixing matrix elements, especially to $\widetilde V_{\alpha i}^{}$ (for $\alpha i = \mu 3, \mu 4,\tau 3,\tau 4$) when $A$ is big enough.  
\subsection{(IMO, $\overline\nu_{3+1}^{}$)}
Similarly, let us discuss the case of an antineutrino beam with inverted mass ordering in the (3+1) flavor mixing scheme.
The evolutions of $\widetilde \Delta_{ij}^{}$ (for $ij =21,31,41$) and $|\widetilde V_{\alpha i}^{}|^2$ (for $\alpha=e,\mu,\tau,s$
and $i=1,2,3,4$) with
the matter effect parameter $A$ in this case are illustrated in the lower right panel of Fig. 7 and  Fig. 12,
respectively. Expanding Eq. (43) in terms of $\Delta_{21}^{}/A$, $\Delta_{31}^{}/A$, $\Delta_{41}^{}/A$ and
$\epsilon$, we arrive at
%%%%%%%%%%%%%%%%%% figure 11 %%%%%%%%%%%%%%%%%%%%%%%
\begin{figure}[h]
	\centering{
		\includegraphics[width=16cm]{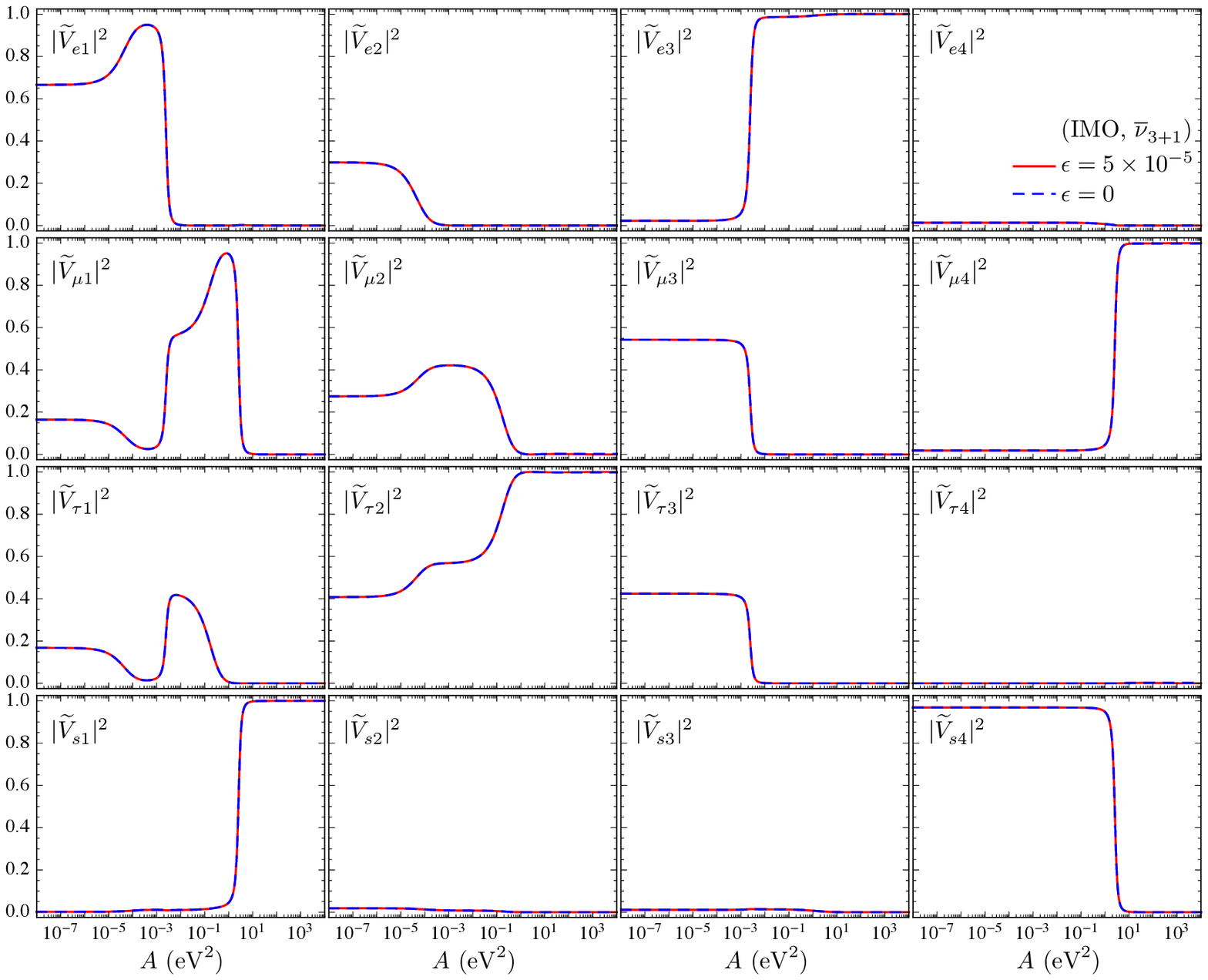}}
	\caption{In the $(3+1)$ mixing scheme, the illustration of  how $|\widetilde V_{\alpha i}^{}|^2$
		(for $\alpha=e,\mu,\tau,s~ {\rm and} ~i=1,2,3,4$) evolve with the matter effect parameter $A$ in the
		case $({\rm IMO}, \overline\nu_{3+1}^{})$ with or without radiative corrections, where we typically take the best-fit values of 
		$(\theta_{12}^{}, \theta_{13}^{}, \theta_{23}^{}, \delta,\Delta_{21}^{}, \Delta_{31}^{} )$ \cite{Esteban:2018azc}, and for the active-sterile neutrino mixing part  $(\theta_{14}^{}, \theta_{24}^{},\theta_{34}^{})=(6.66^{\circ}, 7.81^{\circ}, 0)$, $\delta_{14}^{}=\delta_{24}^{}=\delta_{34}^{}=0$, $\Delta_{41}^{}=1.32~ {\rm eV^2}$ \cite{Diaz:2019fwt}.}
\end{figure}
%%%%%%%%%%%%%%%%%%%%%%%%%%%%%%%%%%%%%%%%%%%%%%%%%%%
\begin{eqnarray}
\widetilde\Delta_{21}^{} &\simeq& \frac{A}{2}  -\frac{1}{2} \left[ A\epsilon + \xi_{s}^{} -
\left(1-|V_{e2}^{}|^2 -3  |V_{s2}^{}|^2\right) \Delta_{21}^{}
- \left( 1-|V_{e3}^{}|^2 - 3 |V_{s3}^{}|^2\right) \Delta_{31}^{} \right.\nonumber\\
&&\left.
- \left( 1- |V_{e4}^{}|^2 - 3  |V_{s4}^{}|^2\right) \Delta_{41}^{}\right] \nonumber\;,\\ 
\widetilde\Delta_{31}^{} &\simeq& -\frac{A}{2}  
+ \left(|V_{e2}^{}|^2 - |V_{s2}^{}|^2\right) \Delta_{21}^{}
+\left(|V_{e3}^{}|^2 -  |V_{s3}^{}|^2\right) \Delta_{31}^{} 
+ \left( |V_{e4}^{}|^2 -  |V_{s4}^{}|^2\right) \Delta_{41}^{} \nonumber\;,\\
\widetilde\Delta_{41}^{} &\simeq& \frac{A}{2} - \frac{1}{2} \left[  A \epsilon - \xi_{s}^{} 
- \left(1 -|V_{e2}^{}|^2 - 3|V_{s2}^{}|^2\right) \Delta_{21}^{}
-\left(1 - |V_{e3}^{}|^2 - 3 |V_{s3}^{}|^2\right) \Delta_{31}^{} \right.\nonumber\\&&\left.
-\left(1 -|V_{e4}^{}|^2 - 3 |V_{s4}^{}|^2\right) \Delta_{41}^{}\right]\;,
%                    (70)
\end{eqnarray}
with $\xi_{s}^{} $ being defined in Eq. (66). According to Fig. 7 and Eq. (70), we always have 
$\widetilde \Delta_{31}^{}<0<\widetilde \Delta_{21}^{}<\widetilde \Delta_{41}^{}$ and in the
$A \to \infty$ limit, $|\widetilde \Delta_{31}^{}|\simeq \widetilde \Delta_{41}^{}>\widetilde \Delta_{21}^{}$ holds.
Similarly, the analytical approximations of $|\widetilde V_{\alpha i}^{}|^2$
read as
\begin{eqnarray}
|\widetilde{V}_{e1}^{}|^2   \simeq  |\widetilde{V}_{e2}^{}|^2  &\simeq & |\widetilde{V}_{e4}^{}|^2 \simeq
0 \; , 
\quad |\widetilde{V}_{s2}^{}|^2   \simeq  |\widetilde{V}_{s3}^{}|^2  \simeq  |\widetilde{V}_{s4}^{}|^2 \simeq
0 \; , 
\quad |\widetilde{V}_{e3}^{}|^2 \simeq |\widetilde{V}_{s1}^{}|^2 \simeq 1 \; ,
\nonumber \\
|\widetilde{V}_{\mu 2}^{}|^2   \simeq |\widetilde{V}_{\tau 4}^{}|^2 &\simeq&
\frac{1}{2} +\frac{1}{2 \xi_{s}^{}}\left[-A \epsilon+ \Delta_{21}^{} \left(
|V_{\tau 2}^{}|^2 - |V_{\mu 2}^{}|^2 \right) +
\Delta_{31}^{} \left(|V_{\tau 3}^{}|^2 - |V_{\mu 3}^{}|^2
\right) \nonumber\right.\\&&\left.+
\Delta_{41}^{} \left(|V_{\tau 4}^{}|^2 - |V_{\mu 4}^{}|^2
\right)\right] \; ,
\nonumber\\
|\widetilde{V}_{\mu 4}^{}|^2   \simeq  |\widetilde{V}_{\tau 2}^{}|^2 &\simeq&
\frac{1}{2} - \frac{1}{2 \xi_{s}^{}}\left[-A \epsilon+ \Delta_{21}^{} \left(
|V_{\tau 2}^{}|^2 - |V_{\mu 2}^{}|^2 \right) +
\Delta_{31}^{} \left(|V_{\tau 3}^{}|^2 - |V_{\mu 3}^{}|^2
\right) \right.\nonumber\\&&\left.+
\Delta_{41}^{} \left(|V_{\tau 4}^{}|^2 - |V_{\mu 4}^{}|^2
\right)\right] \; ,\nonumber\\
|\widetilde{V}_{\mu 1}^{}|^2   \simeq  |\widetilde{V}_{\mu 3}^{}|^2 & \simeq&  |\widetilde{V}_{\tau 1}^{}|^2  \simeq  |\widetilde{V}_{\tau 3}^{}|^2\simeq 0 \;,
%           (71)
\end{eqnarray}
which implies
\begin{eqnarray}
\left. \widetilde{V}\right|_{A\gg\Delta_{41}^{}} \simeq
\begin{pmatrix} 0 & 0 & 1 & 0\cr 
0 &\cos\theta_s^{}&0& \sin\theta_s^{} \cr
0&-\sin\theta_s^{}&0& \cos\theta_s^{}  \cr
1&0 &0&0
\end{pmatrix} \; ,
%     (72)
\end{eqnarray}
where $\theta_s^{} \in [0, \pi/2]$ and $\tan^2\theta_s^{}=|\widetilde V_{\mu 4}^{}|^2/ |\widetilde V_{\mu 2}^{}|^2 $ with $|\widetilde V_{\mu 2}^{}|^2$ and  $|\widetilde V_{\mu 4}^{}|^2$ being expressed in Eq. (71).
One can directly write out the corresponding approximations of neutrino oscillation probability by replacing $\widetilde \Delta_{21}^{}$  in Eq. (59) with $\widetilde \Delta_{42}^{} \simeq \xi_s^{}$ from Eq. (70).
If the smaller terms of $A\epsilon$ in Eq. (71) is negligible,  we obtain
\begin{eqnarray}
|\widetilde V_{\mu 2}^{}|^2 &\simeq& |\widetilde V_{\tau 4}^{}|^2 \simeq \frac{|V_{\tau 4}^{}|^2}{ |V_{\mu 4}^{}|^2 + |V_{\tau 4}^{}|^2 }
\;, \nonumber\\    
|\widetilde V_{\mu 4}^{}|^2 &\simeq&  |\widetilde V_{\tau 2}^{}|^2 \simeq\frac{|V_{\mu 4}^{}|^2}{ |V_{\mu 4}^{}|^2 + |V_{\tau 4}^{}|^2 }
%             (73)
\end{eqnarray} 
after throwing out the smaller terms of $\Delta_{21}^{}$ and $\Delta_{31}^{}$. By inputting
$\theta_{34}^{} =0$ taken in Fig. 12, Eq. (73) turns out to be $|\widetilde V_{\mu 2}^{}|^2 \simeq |\widetilde V_{\tau 4}^{}|^2 \simeq 0$ and $|\widetilde V_{\mu 4}^{}|^2 \simeq |\widetilde V_{\tau 2}^{}|^2 \simeq 1$.
In the $A \to \infty$ limit, we get $|\widetilde V_{\mu 2}^{}|^2 \simeq |\widetilde V_{\tau 4}^{}|^2 \simeq 0$ and $|\widetilde V_{\mu 4}^{}|^2 \simeq |\widetilde V_{\tau 2}^{}|^2 \simeq 1$ no matter which value $\theta_{34}^{}$ takes. 
Thus no neutrino oscillations between the four flavors will happen with $\theta_s^{}\simeq \arctan(|\widetilde V_{\mu 4}^{}|/|\widetilde V_{\mu 2}^{}|)  \simeq \pi/2 $.
Note that the neutrino flavor mixing with radiative corrections can be very different from the case without radiative corrections if $\theta_{34}^{}$ is non-zero.
 
\section{Summary}
With the coming of the precision measurement era of neutrino physics, 
we are committed to digging the underlying physics behind the lepton flavor mixing \cite{Xing:2019vks}
and on the other hand to conducting cosmological and astronomical researches with neutrinos being
a good probe. As preliminarily discussed in Ref. \cite{Luo:2019efb}, it is possible to explore the density and size of a hidden compact object in the universe by observing its effects on the neutrino flavor mixing. In this paper, we point out that radiative corrections to the matter potentials can significantly
affect the neutrino flavor mixing in dense matter. Considering the standard three-flavor mixing scheme with radiative corrections, we derive the exact expressions of the effective neutrino mass-squared differences $\widetilde \Delta_{ij}$, the moduli square of the nine lepton flavor mixing matrix elements $|\widetilde U_{\alpha i}^{}|^2$, the vector sides of the Dirac leptonic unitarity triangles $\widetilde U_{\alpha i}^{}\widetilde U_{\beta i}^{*}$ in a medium. From these exact formulas, the neutrino flavor mixing in dense matter are numerically
and analytically discussed. Different from the fixed value of $|\widetilde U_{\alpha i}^{}|^2$ in dense matter in
the case without radiative corrections, $|\widetilde U_{\alpha i}^{}|^2$ can be very sensitive to 
the value of $A$ and trivially approach 0 or 1 in the $A\to\infty$ limit if  radiative corrections are taken into
account. When it comes to the $(3+1)$ flavor mixing scheme, the neutrino flavor mixing will be very different from the standard three-flavor scheme if A is big enough but not infinite. However it is meaningless
to discuss the lepton flavor mixing in both schemes in the $A\to\infty$ limit.
\section{Acknowledgements}

We would like to thank Prof. Zhi-zhong Xing and Prof. Shun Zhou for their inspirations on this topic.
We are also indebted to Di Zhang for useful discussions.
This work is supported partly by the National Natural Science Foundation of China under grant No. 11835005 and No.  11947227.
%\newpage

\end{document}